\documentclass[iop,twocolumn,twocolappendix]{emulateapj}
\usepackage{natbib}
\usepackage{graphicx}
\usepackage{epsfig}
\usepackage{textcomp}
\usepackage{subfigure}
\usepackage{booktabs}
\usepackage{color}
\usepackage{soul}
\usepackage{floatrow}
\usepackage{hyperref}

\slugcomment{}
\shorttitle{PAH emission toward the Galactic bulge}
\shortauthors{M. J. Shannon et al.}

\begin{document}
\newcommand{\mt}{\textmu m~}
\newcommand{\m}{\textmu m}
\newcommand{\ha}{H$\alpha$~}
\newcommand{\HII}{\mbox{H\,\textsc{ii}}}%
\newcommand*{\thead}[1]{\multicolumn{1}{c}{#1}}
\renewcommand{\deg}{\ensuremath{^{\circ}}}
\newcommand{\td}{\textdegree}

\title{Polycyclic aromatic hydrocarbon emission toward the Galactic bulge}
\author{M. J. Shannon\altaffilmark{$\dagger$,1}, E. Peeters\altaffilmark{1,2}, J. Cami\altaffilmark{1,2}, J.~A.~D.~L. Blommaert\altaffilmark{3}}

\altaffiltext{$\dagger$}{Contact: matthew.j.shannon@gmail.com}
\altaffiltext{1}{Department of Physics and Astronomy and Centre for Planetary Science and Exploration (CPSX), The University of Western Ontario, London, ON N6A 3K7, Canada}
\altaffiltext{2}{SETI Institute, 189 Bernardo Avenue, Suite 100, Mountain View, CA 94043, USA}
\altaffiltext{3}{Astronomy and Astrophysics Research Group, Department of Physics and Astrophysics, Vrije Universiteit Brussel, Pleinlaan 2, 1050 Brussels, Belgium}

\begin{abstract}
We examine polycyclic aromatic hydrocarbon (PAH), dust and atomic/molecular emission toward the Galactic bulge using \textit{Spitzer} Space Telescope observations of four fields: C32, C35, OGLE and NGC 6522. These fields are approximately centered on (l, b) = (0.0\td, 1.0\td), (0.0\td, -1.0\td), (0.4\td, -2.1\td) and (1.0\td, -3.8\td), respectively. Far-infrared photometric observations complement the Spitzer/IRS spectroscopic data and are used to construct spectral energy distributions. We find that the dust and PAH emission are exceptionally similar between C32 and C35 overall, in part explained due to their locations---they reside on or near boundaries of a 7 Myr-old Galactic outflow event and are partly shock-heated. Within the C32 and C35 fields, we identify a region of elevated \ha emission that is coincident with elevated fine-structure and [O~\textsc{iv}] line emission and weak PAH feature strengths. We are likely tracing a transition zone of the outflow into the nascent environment. PAH abundances in these fields are slightly depressed relative to typical ISM values. In the OGLE and NGC 6522 fields, we observe weak features on a continuum dominated by zodiacal dust. SED fitting indicates that thermal dust grains in C32 and C35 have comparable temperatures to those of diffuse, high-latitude cirrus clouds. Little variability is detected in the PAH properties between C32 and C35,  indicating that a stable population of PAHs dominates the overall spectral appearance. In fact, their PAH features are exceptionally similar to that of the M82 superwind, emphasizing that we are probing a local Galactic wind environment.
\end{abstract}


\section{Introduction}
\label{sec:intro}

Polycyclic aromatic hydrocarbons (PAHs) are a highly abundant family of astronomical molecules that produce dominant infrared (IR) emission features in the 3-20 \mt spectral range, primarily at 3.3, 6.2, 7.7, 8.6, 11.2, 12.7 and 16.4 \mt (e.g., \citealt{leger1984,allamandola1985,allamandola1989,peeters2002}). They are common in the interstellar medium and in carbon-rich circumstellar environments, and as such most lines of sight are littered with PAH emission features (for a review see \citealt{tielens2008} and references therein). The PAH emission bands are highly variable, changing in band profile, absolute strength and \textit{relative} strength from environment to environment(e.g., \citealt{hony2001,peeters2002,galliano2008b}). These variations have been linked to the local physical conditions via (e.g.,) radiation field strength, electron density and gas temperature, showing diagnostic potential in using PAHs as probes of local conditions (e.g., \citealt{sloan2007,galliano2008b,boersma2013,stock2016}).

PAHs are excited by ultraviolet and visible photons, glowing brightly near regions of ongoing star formation. Conversely, PAH emission is generally weak in sight-lines with no illuminating source. \citet{golriz2014} studied AGB stars in the Galactic bulge and identified PAH emission in background positions of their observations. The Galactic bulge consists mostly of an old population of stars (10 $\pm$ 2.5 Gyr; \citealt{ortolani1995,zoccali2003}). Also, an intermediate age (1-3 Gyr) stellar population exists as evidenced by Mira variables which have evolved from a population of 1.5-2 M$_\sun$ stars \citep{groenewegen2005,blommaert2006}. Many of the bulge stars are in the asymptotic giant branch (AGB) phase, or at the tip of the red giant branch phase (e.g. \citealt{omont1999,ojha2003}). We emphasize that we are only examining background (off-star) positions--altogether, the source of PAH excitation in this environment is not immediately obvious, particularly in conjunction with the strong fine-structure lines observed towards these positions.

As such, we aim to characterize the PAH, dust and fine-structure line emission toward the Galactic bulge, and correlate these properties to the local physical conditions. We present a mid- and far-IR study of emission towards the Galactic bulge in four fields. One field of observations (C32) is on the edge of the Galactic center lobe (GCL), a several hundred parsec feature slightly north of the Galactic plane. It was first identified by \citet{sofue1984} in radio continuum emission, who found that it spans roughly 185 pc $\times$ 210 pc between $l=359.2\deg - 0.2\deg$ and $b=0.2\deg - 1.2\deg$. We use known properties of the GCL to interpret the PAH, dust and fine structure line emission towards C32. Another field (C35) appears to reside near the edge of a complementary lobe south of the Galactic plane. Fields OGLE and NGC 6522 are further south of the plane, but generally have limited survey coverage in comparison. We thus focus especially on C32 and C35 in this analysis.

We detail our observations and data reduction methods in Sec.~\ref{sec:obs} and accompanying data analysis in Sec.~\ref{sec:inventory}. Results are presented in Sec.~\ref{sec:results} and we discuss relevant implications in Sec.~\ref{sec:discussion}. Lastly, a brief summary of this work is presented in Sec.~\ref{sec:conclusion}.

\section{Observations and Data Reduction}
\label{sec:obs}

\subsection{Target selection and observations}

\begin{figure}
	\centering
    \includegraphics[width=1\linewidth, clip=true, trim=4.1cm 0.7cm 5.8cm 1.4cm]{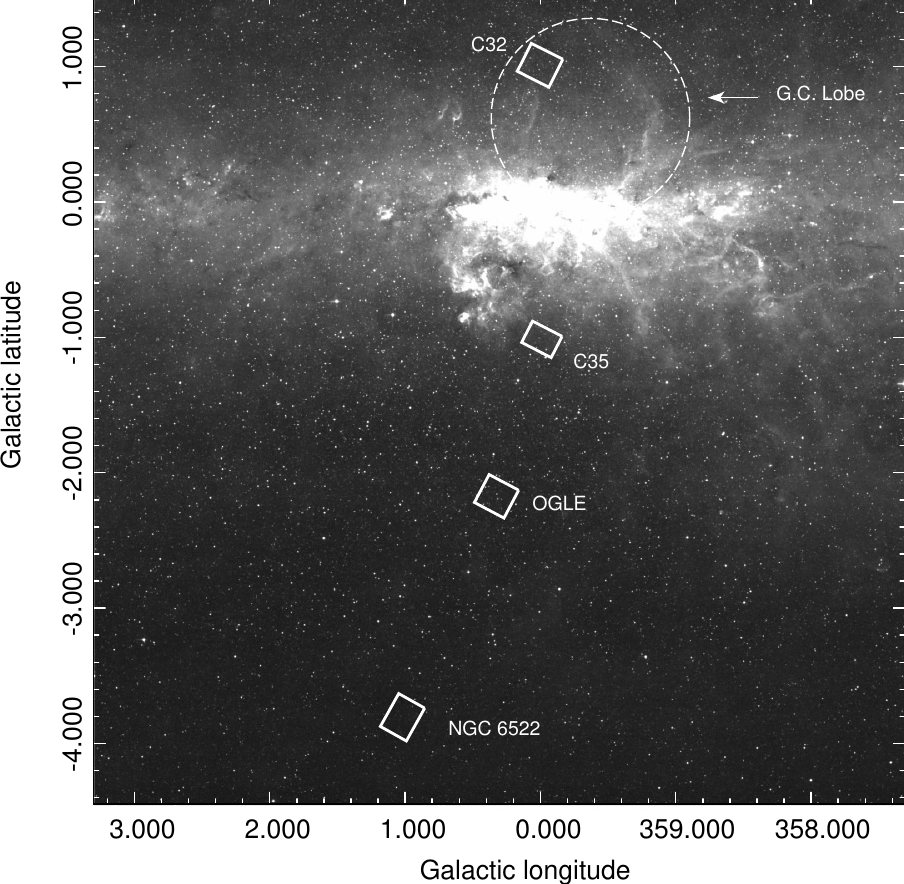}
	\caption{
	An overview of our \textit{Spitzer}/IRS fields (labeled rectangles) overlaid on the 8.3 \mt band from the \textit{Midcourse Space Experiment} (MSX; \citealt{mill1994}) survey of the Galactic plane \citep{price2001}. The field of C32 is coincident with the boundary of the Galactic center lobe, which is the large wispy arc (located within the dashed circle, drawn to help guide the eye). Coordinates are presented in degrees.
	}
	\label{fig:halpha}
\end{figure}

\citet{golriz2014} reported strong IR background emission (including prevalent PAH features) towards the Galactic bulge in four fields (C32, C35, OGLE, NGC 6522), each of which contains multiple pointings. These fields lie at different projected distances to the Galactic bulge (Fig.~\ref{fig:halpha}), notably with C32 and C35 being diametrically opposed across the Galactic center at (l, b) = (0.0\td, 1.0\td), (0.0\td, -1.0\td), respectively. The OGLE and NGC 6522 fields are further south of the Galactic plane, near (0.4\td, -2.1\td) and (1.0\td, -3.8\td), respectively. Note that NGC 6522 is in Baade's Window.

\begin{figure}
\begin{center}
    \includegraphics[width=1\linewidth, clip=true, trim=1cm 1.2cm 2cm 0.8cm] {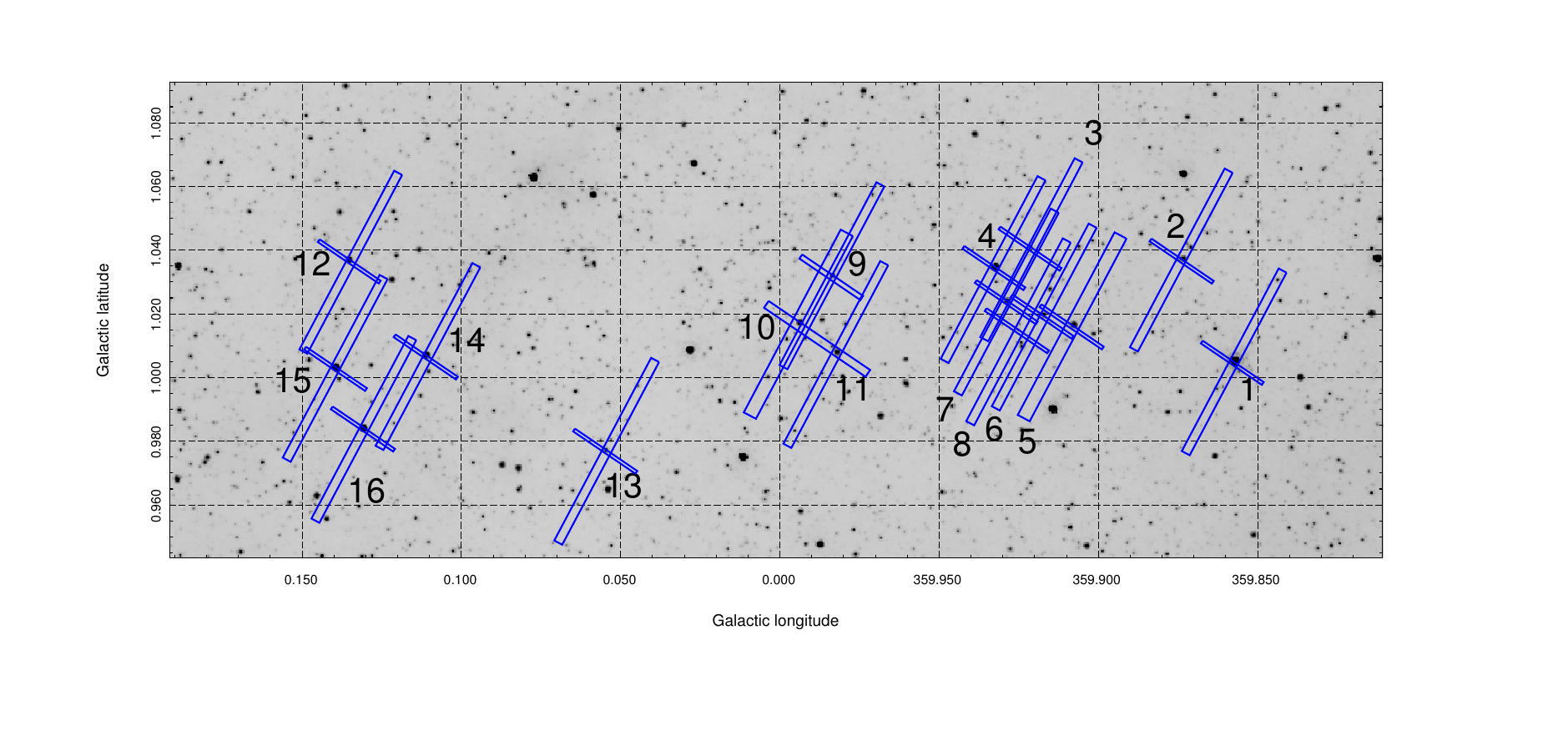}
    \includegraphics[width=0.98\linewidth, clip=true, trim=2.6cm 1.3cm 4cm 1.8cm] {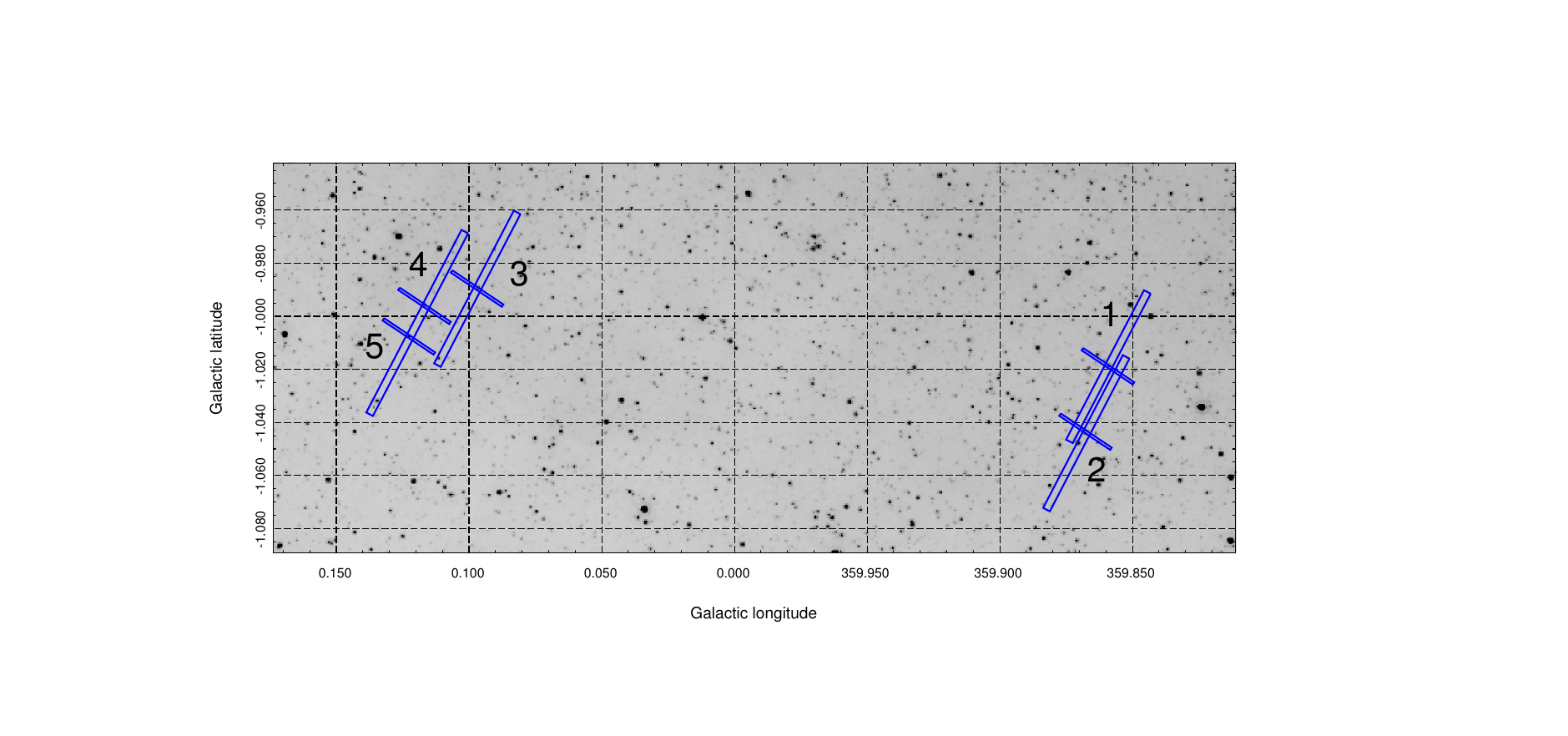}
\caption{The IRS apertures for C32 (\textit{top}) and C35 (\textit{bottom}) are overlaid on an Infrared Array Camera (IRAC; \citealt{fazio2004}) 8 \mt image. The numbered labels correspond to the overlapping SL and LL apertures (the short and long blue rectangles, respectively), e.g. C32-1 (c.f. Table~\ref{table:obs}).}
\label{fig:irac_c32c35}
\end{center}
\end{figure}

The spectroscopic observations were acquired with the Infrared Spectrograph (IRS; \citealt{houck2004}) on the \textit{Spitzer} Space Telescope \citep{werner2004}. These data were obtained from the NASA/IPAC \textit{Spitzer} Heritage Archive.\footnote{\url{http://sha.ipac.caltech.edu/applications/Spitzer/SHA/}} The low-resolution ($R\sim100$) data span approximately 5-40 \m, using the short low (SL) and long low (LL) modules. These data were previously examined by \citet{golriz2014} for the purpose of studying bulge AGB stars, which lie at the center of each pointing. However, we are only interested in the off-source emission. A summary of our observations is presented in Table~\ref{table:obs}, which is based on the sample of \citet{golriz2014}, their Table 1. Each field consists of multiple spectral maps, with corresponding unique identifiers, as illustrated in Figs.~\ref{fig:irac_c32c35},~\ref{fig:irac_ogle} and~\ref{fig:irac_ngc6522}, for the fields of C32, C35, OGLE and NGC 6522, respectively. Our sample contains a total of 47 separate pointings across these four fields.

\begin{table*}
\begin{center}
\resizebox{0.85\linewidth}{!}{
\begin{tabular}{lcccrrc}
\toprule
\toprule
ID & Object$^a$ & RA (J2000) & Dec. (J2000) & l (deg.) & b (deg.) & AOR key$^b$ \\
\midrule
C32-1	&	J174117.5-282957	&	17:41:17.50	&	-28:29:57.50	&	359.874	&	1.037	&	10421504	\\
C32-2	&	J174122.7-283146	&	17:41:22.70	&	-28:31:47.00	&	359.858	&	1.005	&	10421504	\\
C32-3	&	J174123.6-282723	&	17:41:23.56	&	-28:27:24.20	&	359.922	&	1.041	&	10422784	\\
C32-4	&	J174126.6-282702	&	17:41:26.60	&	-28:27:02.20	&	359.933	&	1.034	&	10421504	\\
C32-5	&	J174127.3-282851	&	17:41:27.26	&	-28:28:52.10	&	359.908	&	1.016	&	10421504	\\
C32-6	&	J174127.9-282816	&	17:41:27.88	&	-28:28:17.10	&	359.918	&	1.019	&	10421504	\\
C32-7	&	J174128.5-282733	&	17:41:28.51	&	-28:27:33.80	&	359.929	&	1.024	&	10421504	\\
C32-8	&	J174130.2-282801	&	17:41:30.15	&	-28:28:01.30	&	359.926	&	1.015	&	10422784	\\
C32-9	&	J174134.6-282431	&	17:41:34.60	&	-28:24:31.40	&	359.984	&	1.032	&	10421504	\\
C32-10	&	J174139.5-282428	&	17:41:39.48	&	-28:24:28.20	&	359.994	&	1.017	&	10421504	\\
C32-11	&	J174140.0-282521	&	17:41:39.94	&	-28:25:21.20	&	359.982	&	1.008	&	10421504	\\
C32-12	&	J174155.3-281638	&	17:41:55.27	&	-28:16:38.70	&	0.135	&	1.037	&	10421504	\\
C32-13	&	J174157.6-282237	&	17:41:57.53	&	-28:22:37.70	&	0.055	&	0.977	&	10421504	\\
C32-14	&	J174158.8-281849	&	17:41:58.73	&	-28:18:49.20	&	0.111	&	1.007	&	10421504	\\
C32-15	&	J174203.7-281729	&	17:42:03.69	&	-28:17:29.90	&	0.139	&	1.003	&	10421504	\\
C32-16	&	J174206.85-281832	&	17:42:06.86	&	-28:18:32.40	&	0.131	&	0.984	&	10421504	\\
C35-1	&	J174917.0-293502	&	17:49:16.96	&	-29:35:02.70	&	359.859	&	-1.019	&	10421248	\\
C35-2	&	J174924.1-293522	&	17:49:23.99	&	-29:35:22.20	&	359.868	&	-1.044	&	10421248	\\
C35-3	&	J174943.7-292154	&	17:49:43.65	&	-29:21:54.50	&	0.097	&	-0.989	&	10421248	\\
C35-4	&	J174948.1-292104	&	17:49:48.05	&	-29:21:04.80	&	0.117	&	-0.996	&	10421248	\\
C35-5	&	J174951.7-292108	&	17:49:51.65	&	-29:21:08.70	&	0.122	&	-1.008	&	10421248	\\
OGLE-1	&	J175432.0-295326	&	17:54:31.94	&	-29:53:26.50	&	0.176	&	-2.156	&	10422528	\\
OGLE-2	&	J175456.8-294157	&	17:54:56.80	&	-29:41:57.40	&	0.387	&	-2.137	&	10422528	\\
OGLE-3	&	J175459.0-294701	&	17:54:58.98	&	-29:47:01.40	&	0.318	&	-2.186	&	10422528	\\
OGLE-4	&	J175511.9-294027	&	17:55:11.90	&	-29:40:27.80	&	0.436	&	-2.171	&	10423040	\\
OGLE-5	&	J175515.4-294122	&	17:55:15.41	&	-29:41:22.80	&	0.429	&	-2.190	&	10423040	\\
OGLE-6	&	J175517.0-294131	&	17:55:16.97	&	-29:41:31.90	&	0.430	&	-2.196	&	10423040	\\
OGLE-7	&	J175521.7-293912	&	17:55:21.70	&	-29:39:13.00	&	0.472	&	-2.192	&	10423040	\\
NGC 6522-1	&	J180234.8-295958	&	18:02:34.78	&	-29:59:58.90	&	0.950	&	-3.722	&	10421760	\\
NGC 6522-2	&	J180238.8-295954	&	18:02:38.72	&	-29:59:54.60	&	0.958	&	-3.734	&	10421760	\\
NGC 6522-3	&	J180248.9-295430	&	18:02:48.90	&	-29:54:31.00	&	1.054	&	-3.722	&	10422016	\\
NGC 6522-4	&	J180249.5-295853	&	18:02:49.44	&	-29:58:53.40	&	0.992	&	-3.759	&	10422272	\\
NGC 6522-5	&	J180259.6-300254	&	18:02:59.51	&	-30:02:54.30	&	0.951	&	-3.824	&	10421760	\\
NGC 6522-6	&	J180301.6-300001	&	18:03:01.60	&	-30:00:01.10	&	0.997	&	-3.807	&	10422272	\\
NGC 6522-7	&	J180304.8-295258	&	18:03:04.80	&	-29:52:59.30	&	1.105	&	-3.760	&	10422272	\\
NGC 6522-8	&	J180305.3-295515	&	18:03:05.25	&	-29:55:15.90	&	1.072	&	-3.780	&	10421760	\\
NGC 6522-9	&	J180305.4-295527	&	18:03:05.33	&	-29:55:27.80	&	1.070	&	-3.782	&	10422016	\\
NGC 6522-10	&	J180308.2-295747	&	18:03:08.11	&	-29:57:48.00	&	1.040	&	-3.809	&	10422016	\\
NGC 6522-11	&	J180308.6-300526	&	18:03:08.52	&	-30:05:26.50	&	0.930	&	-3.873	&	10421760	\\
NGC 6522-12	&	J180308.7-295220	&	18:03:08.69	&	-29:52:20.40	&	1.121	&	-3.767	&	10421760	\\
NGC 6522-13	&	J180311.5-295747	&	18:03:11.47	&	-29:57:47.20	&	1.047	&	-3.820	&	10421760	\\
NGC 6522-14	&	J180313.9-295621	&	18:03:13.88	&	-29:56:20.90	&	1.072	&	-3.816	&	10422016	\\
NGC 6522-15	&	J180316.1-295538	&	18:03:15.99	&	-29:55:38.30	&	1.086	&	-3.817	&	10422272	\\
NGC 6522-16	&	J180323.9-295410	&	18:03:23.84	&	-29:54:10.70	&	1.121	&	-3.830	&	10422016	\\
NGC 6522-17	&	J180328.4-295545	&	18:03:28.36	&	-29:55:45.40	&	1.106	&	-3.856	&	10421760	\\
NGC 6522-18	&	J180333.3-295911	&	18:03:33.26	&	-29:59:11.50	&	1.065	&	-3.900	&	10422016	\\
NGC 6522-19	&	J180334.1-295958	&	18:03:34.07	&	-29:59:58.80	&	1.055	&	-3.909	&	10421760	\\
\bottomrule
\end{tabular}
}
\end{center}

This table is adapted from \citet{golriz2014}, their Table 1. The coordinates are the central positions of the slit. $^a$References: \citet{omont2003,ojha2003,blommaert2006}. $^b$The AOR key uniquely identifies \textit{Spitzer} Space Telescope observations.
\caption{Spitzer/IRS observations}
\label{table:obs}
\end{table*}

Photometric observations between 12 and 500 \mt are included from three sources to augment the mid-IR spectroscopy. First, we include data from the \textit{Herschel} Space Observatory \citep{pilbratt2010} infrared Galactic plane survey (Hi-GAL; \citealt{molinari2010}), which observed the Galactic plane with the Photoconductor Array Camera and Spectrometer (PACS; \citealt{poglitsch2010}) and Spectral and Photometric Imaging REceiver (SPIRE; \citealt{griffin2010}). Second, we include photometric observations taken by the \textit{AKARI} Space Telescope \citep{murakami2007} with its Far-Infrared Surveyor (FIS) instrument \citep{kawada2007}, released as part of the \textit{AKARI} all-sky survey maps \citep{doi2015}. And third, we analyze images obtained by the \textit{Infrared Astronomical Satellite} (\textit{IRAS}), which were released via Improved Reprocessing of the \textit{IRAS} Survey (IRIS; \citealt{iris2005}). The photometric observations are summarized in Table~\ref{table:photometry}. Archival \ha images of the Galactic bulge region are acquired from the Southern \ha Sky Survey Atlas (SHASSA; \citealt{gaustad2001}).

Photometric calibration errors of the Hi-GAL photometric observations have been estimated as 5\% for \textit{Herschel}/PACS and 4\% for \textit{Herschel}/SPIRE \citep{molinari2016}. For the IRIS sample of \textit{IRAS} observations, the errors are approximately 15\%, 18\%, 11\% and 20\% for the 12, 25, 60 and 100 \mt bands, respectively \citep{iris2005}. The \textit{AKARI} photometric errors can be up to 20\%, 30\%, 40\% and 40\% for the 60, 90, 140 and 160 \mt filters \citep{kawada2007}.

\begin{table*}
\begin{center}
\resizebox{1\linewidth}{!}{
\begin{tabular}{l l l l c}
\toprule
\toprule
Observatory            & Instrument   	 & Nominal filters		& Data origin & Data reference \\
\midrule
\textit{Herschel} Space Observatory 	& PACS, SPIRE     & 70, 160, 250, 350, 500 \m   & Hi-GAL       & 1\\
\textit{AKARI}    						& FIS     			& 60, 90, 140, 160 \m       & All-sky survey maps & 2\\
\textit{Infrared Astronomical Satellite} (IRAS)  &    			& 12, 25, 50, 100 \m        & IRIS 	& 3\\
Cerro Tololo Inter-American Observatory & & 656 nm & SHASSA & 4 \\

\bottomrule
\end{tabular}
}
\end{center}
\centering
References: (1) Hi-GAL: The \textit{Herschel} Infrared Galactic Plane Survey \citep{molinari2010}; (2) \textit{AKARI} Far-Infrared All-Sky Survey Maps \citep{doi2015}; (3) IRIS: Improved Reprocessing of the \textit{IRAS} Survey \citep{iris2005}; (4) SHASSA: Southern \ha Sky Survey Atlas \citep{gaustad2001}.
\caption{Photometric observations}
\label{table:photometry}
\end{table*}

\subsection{Data reduction}

The \textit{Spitzer}/IRS data were reduced using the \texttt{CUBISM} tool \citep{smith2007cubism}, beginning with the basic calibrated data processed by the \textit{Spitzer} Science Center\footnote{\url{http://ssc.spitzer.caltech.edu/}} (pipeline version S18.18). The \texttt{CUBISM} tool performs coaddition and bad pixel cleaning and produces full spectral cubes. Since we are only interested in the background (off-star) emission, no further reduction was performed (apart from identifying additional cosmic ray spikes or bad pixels). The slits were rebinned with a $2\times 2$-pixel aperture to match the point-spread function of IRS to sample independent pixels. Additional details of this approach are described by \citet{peeters2012}. The photometric data were retrieved fully processed, thus no additional reduction was necessary.

\subsection{Aperture overlap}
\label{sec:apertures}

The IRS/SL and IRS/LL apertures are oriented approximately perpendicular to one another, such that the intended astronomical target lies at their intersection. Since we are interested in the background emission in the IRS observations, i.e., not the point of intersection, we cannot measure the 5-14 \mt and 14-38 \mt spectra at the same spatial position, which introduces an offset. After combining nod positions, each SL aperture in our observations typically spans 83\arcsec, while the LL apertures cover 233\arcsec. The maximum separation between SL and LL pointings is thus approximately 118\arcsec, with mean separations near 59\arcsec. We analyzed our spectra in two ways: first, we measured the median spectra in each module (avoiding the stellar emission zone), and stitched the median components together to produce a single full spectrum from 5-38 \m; and second, we analyzed the SL and LL spectra entirely independently. No systematics were identified when comparing the separate spectra to the fully stitched median spectra, so we use the latter to represent the astronomical background emission for each pointing. It should be noted that some of the positions in each field are close enough to other pointings such that (e.g.,) the SL aperture of one position may overlap the LL aperture of another (see Fig.~\ref{fig:irac_c32c35}). However, we are interested in the general behavior of the spectra across the fields, rather than small position-to-position variations. We use apertures that are tightly clustered to verify consistency in our results.

In C32 and C35, calibration offsets between the SL and LL modules and orders are very minor (on average less than 5\%, and reaching a maximum of 10\%), despite the spatial offset between the apertures. In the OGLE and NGC 6522 fields, however, the SL and LL modules are frequently mismatched by 10-20\%, and in one instance as high as 40\%. Because of this, we do not scale the SL and LL modules to each other in OGLE and NGC 6522. In all OGLE and NGC 6522 positions, the individual LL1 and LL2 orders are well matched, but the SL1 and SL2 orders are not (for instance, a factor of 0.30 is typically needed to bring SL2 in line with SL1 for the OGLE and NGC 6522 fields). As such, we exclude the OGLE and NGC 6522 measurements when examining PAH band strength ratios later in the text (Section~\ref{sec:corr} and accompanying figures).

\section{Spectral analysis}
\label{sec:inventory}

\begin{figure*}
	\centering
	\includegraphics[width=0.75\linewidth]{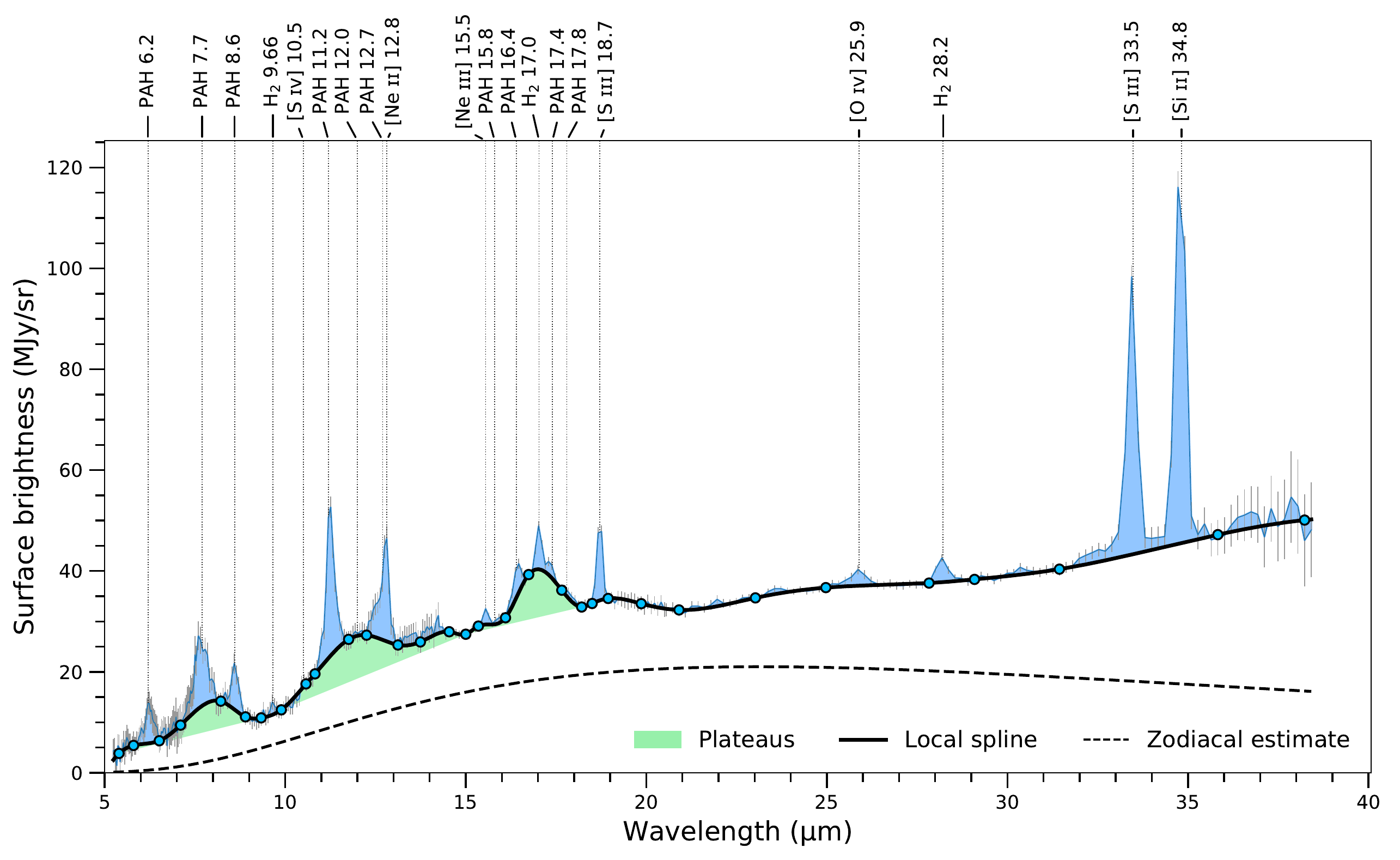}
	\caption{
	The median spectrum of C32-1. The thick black curve is the local spline continuum, while the dashed black line is an estimate of the zodiacal dust emission. Three plateaus are denoted by the green shading. These are determined by measuring the area between the local spline continuum and straight lines fit between the emission at 5, 10, 15 and 18 \m; see Section~\ref{sec:inventory}. The prominent emission features are identified with dotted vertical lines.
    }
	\label{fig:spectrum1}
\end{figure*}

Examining the spectrum for pointing 1 of field C32 (C32-01, for short) in Fig.~\ref{fig:spectrum1}, many PAH emission features are prominent, including bands at 6.2, 7.7, 8.6, 11.2, 12.7, 16.4, 17.4 and 17.8 \m. Weaker PAH emission at 12.0 \m, 15.8 \mt and possibly 14.0 \mt may also be present. A smoothly rising dust continuum is visible, as are plateaus between 5-10 \m, 10-15 \mt and 15-18 \m. Atomic emission lines are also present, including 12.8 \mt [Ne~\textsc{ii}], 15.5 \mt [Ne~\textsc{iii}], 18.7 \mt [S~\textsc{iii}], 25.9 \mt [O~\textsc{iv}], 33.5 \mt [S~\textsc{iii}] and the 34.8 \mt [Si~\textsc{ii}] line, in addition to H$_2$ lines at 9.7, 12.3, 17.0 \mt and 28.2 \m. It is possible in some instances that emission from the 25.99 \mt [Fe~\textsc{ii}] line is blended with the 25.89 \mt [O~\textsc{iv}] line. Our measurements of line centroids suggest however that we are observing [O~\textsc{iv}] the majority of the time, if any blend is present at all. As such, we henceforth assume the emission is from the 25.89 \mt [O~\textsc{iv}] line.

The emission features in each spectrum are isolated from the underlying continuum by fitting a local spline continuum to measure the band and line fluxes (Fig.~\ref{fig:spectrum1}). This spline is anchored at a series of wavelengths where only continuum emission is expected. This is a common approach in the literature for measuring the strengths of the PAH emission features (e.g. \citealt{vankerckhoven2000,hony2001,peeters2002,galliano2008b}). Other possible methods for measuring these bands include fitting Drude profiles (PAHFIT; \citet{smith2007}) or Lorentzian profiles \citep{boulanger1998,smith2007,galliano2008b}. \citet{galliano2008b} showed that the measured PAH fluxes will vary from method to method, but the overall trends and conclusions reached using these methods are consistent.

Once the continuum has been identified and subtracted, the spectra are analyzed in several ways: the PAH features are directly integrated, except for the 11.0 and 12.7 \mt bands, which are blended with the 11.2 \mt emission and the 12.8 \mt [Ne~\textsc{ii}] line, respectively. There may also be 12.3 \mt H$_2$ emission blended with the 12.7 \mt PAH band. The 12.7 \mt PAH band is isolated from the 12.3 \mt H$_2$ and 12.8 \mt [Ne~\textsc{ii}] lines by fitting a template of the 12.7 \mt PAH emission (for details see \citealt{stock2014} and \citealt{shannon2015}). The 11.0 \mt emission is fit with a Gaussian while keeping the shape of the lightly blended 11.2 \mt band  fixed. The other atomic and molecular lines are fit with Gaussians whose widths are fixed to the instrument's spectral resolution.

A common method for estimating the plateau strengths is to fit straight lines between the continuum emission at 5, 10, 15 and 18 \mt \citep{peeters2012,peeters2017}. The difference between this curve and the local spline then defines the plateau regions, which we directly integrate (Fig.~\ref{fig:spectrum1}).

\section{Results}
\label{sec:results}

\subsection{Composite images}

\begin{figure*}
    \centering
    \includegraphics[width=0.7\linewidth, clip=false, trim=0.57cm 0.30cm 0.65cm 0.27cm]{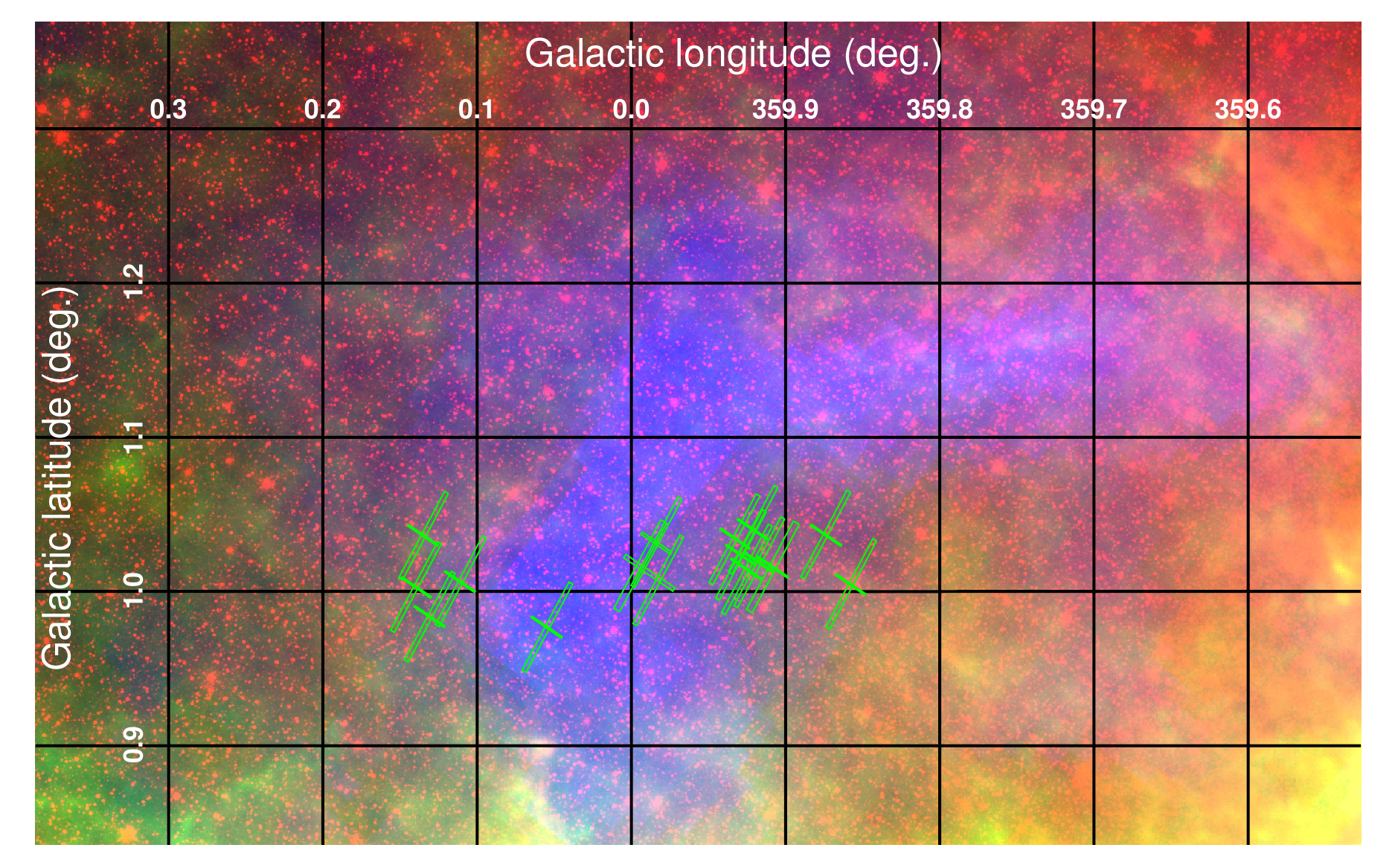}
    \caption{
	A composite image of C32, constructed with images from Spitzer/IRAC 8 \mt (red), Herschel/SPIRE 250 \mt (green) and 656 nm \ha emission (blue). The green rectangles identify the SL and LL apertures (less elongated and more elongated, respectively; c.f. Fig.~\ref{fig:irac_c32c35} and Table~\ref{table:obs} for identifications). An elevated \ha emission zone (or channel) that bisects the IRS apertures is visible.
	}
    \label{fig:rgb_c32}
\end{figure*}

\begin{figure*}
    \centering
    \includegraphics[width=0.7\linewidth]{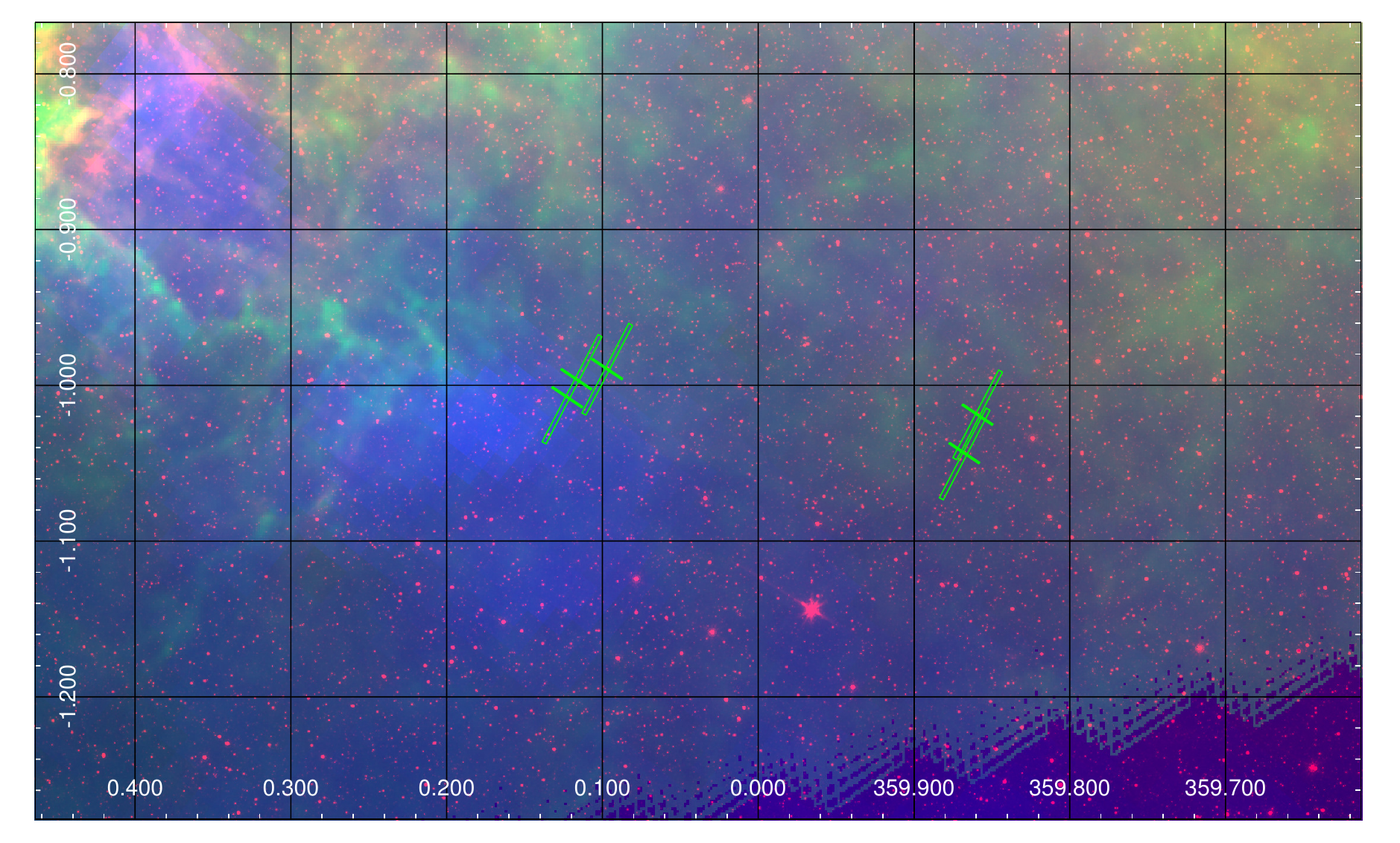}
    \caption{
	A composite image of C35, with Galactic longitude (in degrees) on the x axis and Galactic latitude (in degrees) on the y axis. The image is composed of emission from Spitzer/IRAC 8 \mt (red), Herschel/SPIRE 250 \mt (green) and 656 nm \ha (blue). The leftmost C35 pointings (corresponding to positions C35-3, C35-4 and C35-5) are coincident with an elevated \ha emission region or channel (c.f. Fig~\ref{fig:irac_c32c35}).
    }
    \label{fig:rgb_c35}
\end{figure*}

We present composite 3-color images of the C32 and C35 fields in Figs.~\ref{fig:rgb_c32} and~\ref{fig:rgb_c35}, respectively. Each composite is constructed from \textit{Spitzer}/IRAC (8 \m), \textit{Herschel}/SPIRE (250 \m) and the SHASSA \ha survey (656 nm). In C32, there appears to be a region of elevated \ha emission (or ``channel" hereafter) that bisects the field. The 8 \mt and 250 \mt emission appear to peak on either side of this channel. A similar elevated \ha region/channel is apparent in C35 (Fig.~\ref{fig:rgb_c35}), coincident with positions 3, 4 and 5. There was insufficient coverage to prepare similar figures for the OGLE and NGC 6522 fields.

\subsection{The spectra}

\begin{figure*}
	\centering
	\includegraphics[width=1.0\linewidth]{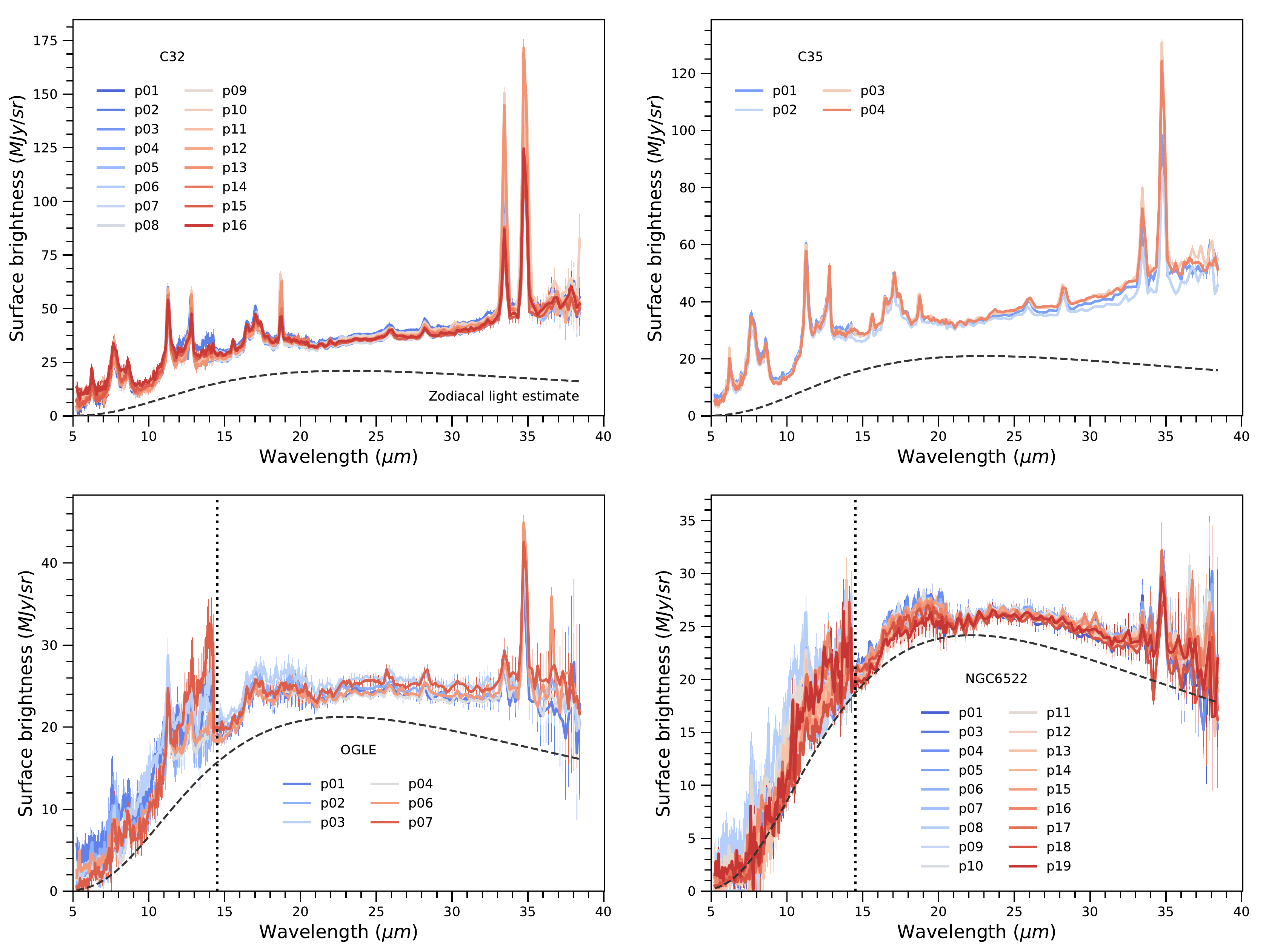}
\caption{Spectra for C32, C35, OGLE and NGC6522 are shown in panels (a), (b), (c) and (d), respectively. These C32 and C35 spectra were created by stitching the emission from the SL and LL modules, which are not necessarily spatially coincident (see Section~\ref{sec:apertures}), and taking a median over each pointing. The OGLE and NGC 6522 SL and LL spectra are not stitched at their overlap (denoted by the vertical dotted line; see Sec.~\ref{sec:apertures}). The colors and inset labels indicate individual positions (c.f., Table~\ref{table:obs}). The dashed black lines below the spectra are estimates for the zodiacal light emission in each field.
 }
\label{fig:meanspec}
\end{figure*}

Median spectra for each position of our four fields (C32, C35, OGLE, NGC 6522) are shown in Fig.~\ref{fig:meanspec}. We have included an estimate for the zodiacal light emission in these spectra using the Spitzer Zodiacal Light Model\footnote{\url{http://irsa.ipac.caltech.edu/data/SPITZER/docs/dataanalysistools/tools/contributed/general/zodiacallight/}}.

C32 displays remarkably similar spectra across the sixteen apertures, both in overall continuum shape and individual emission features (Fig.~\ref{fig:meanspec}a). Deviations in continuum brightness between C32 positions are typically less than 5 MJy/sr on $\sim$20-40 MJy/sr continua. We observe significant contribution from zodiacal light, but it does not dominate the overall continuum of these spectra. The PAH features and plateaus appear to be similar in strength and shape across all positions. Only the atomic fine-structure lines show significant variations, as the 18.7 \mt [S~\textsc{iii}], 33.5 \mt [S~\textsc{iii}] and 34.8 \mt [Si~\textsc{ii}] emission lines vary by a factor of approximately two in peak intensity.  The 25.89 \mt [O~\textsc{iv}] line is also visible, which is typically a tracer of shocked gas \citep{simpson2007}, though it appears to vary little across the field.

The C35 spectra (Fig.~\ref{fig:meanspec}b) are very similar in appearance to the spectra of C32. The only spectral differences within the C35 field are variations in their continuum shape beyond $\sim25$ \m, with position 2 being slightly flatter than the other positions. Fine-structure line variations are observed, similar to that of C32, and the 25.89 [O~\textsc{iv}] line is again clearly present. The contribution from zodiacal light is essentially identical to that observed toward C32---the rising dust continuum at long wavelengths diverges from the zodiacal dust emission.

Spectra for the OGLE field (Fig.~\ref{fig:meanspec}c) display a strong jump between modules SL and LL (near 14.5 \m) in some positions (see Section~\ref{sec:apertures}). Beyond 15 \mt all spectra within the field have comparable continua, with typical surface brightnesses of approximately 24 MJy/sr. The overall shape of the OGLE spectra essentially trace the zodiacal dust emission, in contrast to C32 and C35. PAH features are visible in the OGLE spectra, though some are weak and/or difficult to detect above the noise (e.g., the 12.7 \mt feature in OGLE-6). There is clear 11.2 \mt emission in several of the OGLE positions. The 15-20 PAH emission is distinct from that seen towards the C32 and C35 fields, seemingly extending to 20 \mt instead of 18 \m---such emission has been previously observed by \citet{vankerckhoven2000} and \citet{peeters2006}. The 25.89 [O~\textsc{iv}] emission line is again observed, along with H$_2$, Ne and S emission lines.

Turning to the final field, the emission in NGC 6522 is dominated by zodiacal dust emission (Fig.~\ref{fig:meanspec}d). These spectra are very noisy and show almost no PAH emission features, apart from possibly very weak 7.7 and 11.2 \mt bands in some positions. However, the 15-20 \mt plateau is present and very strong in NGC 6522. The 33.5 \mt [S~\textsc{iii}] and 34.8 \mt [Si~\textsc{ii}] lines are also present at relatively low signal-to-noise. The 25.89 [O~\textsc{iv}] emission line cannot be detected, if it is present at all.

\begin{figure}
	\centering
	\includegraphics[width=1.0\linewidth]{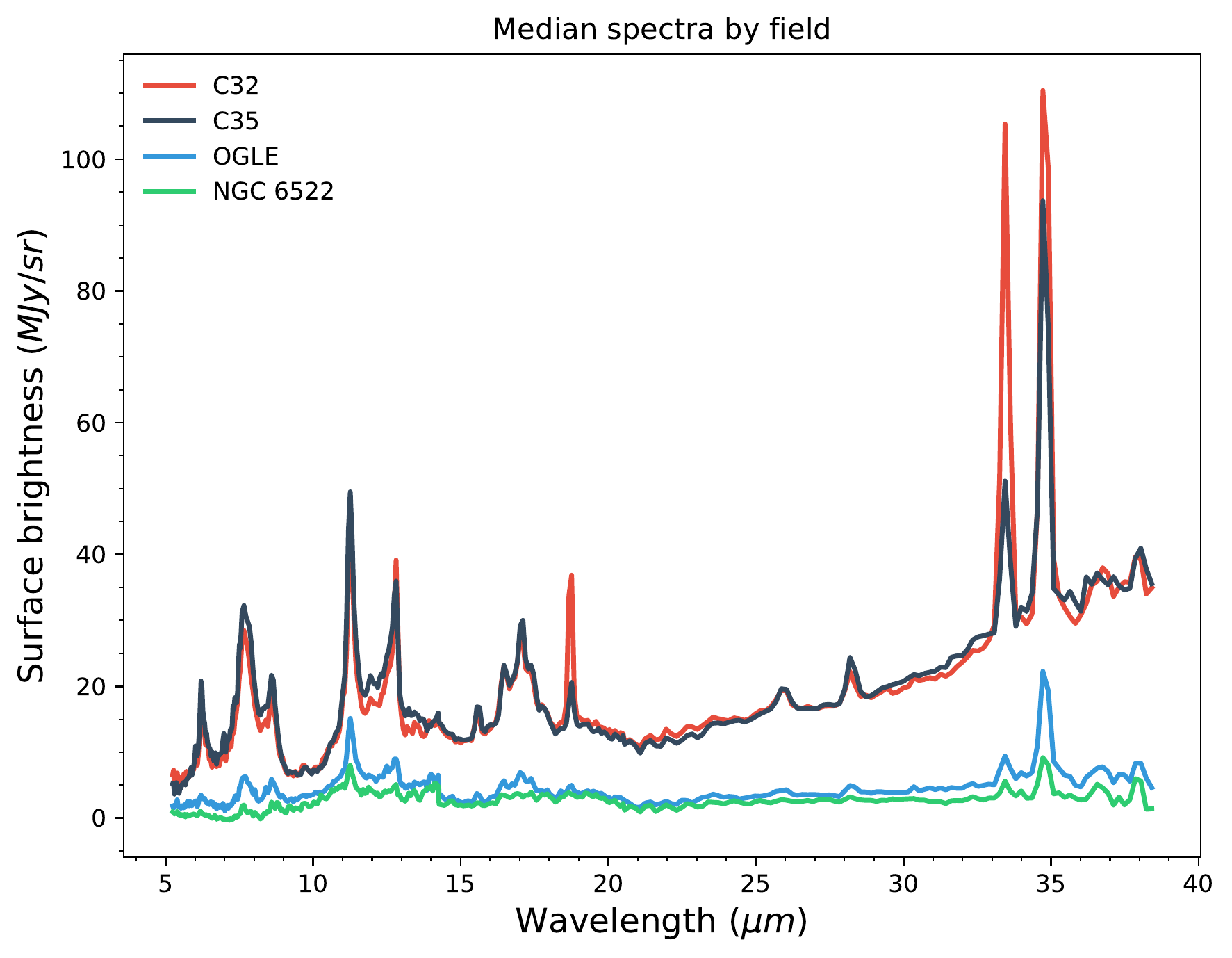}
	\caption{
	A comparison of the median spectra in each field after subtracting the zodiacal dust emission. Surprisingly, the C32 and C35 spectra are exceedingly similar in all but the atomic fine-structure lines. The continua of OGLE and NGC 6522 have essentially disappeared, showing that they were dominated by the zodiacal light.
    }
	\label{fig:specall4}
\end{figure}

We compare all four fields, overlaid, in Fig.~\ref{fig:specall4}, with a single median spectrum being constructed from all positions in each field. The zodiacal dust emission has been removed from these data. The C32 and C35 spectra are extremely similar, differing only in the strength of their atomic fine-structure lines and possibly the continuum near 12-14 \m. The 18.71 and 33.48 \mt [S~\textsc{iii}] lines and 34.82 \mt [Si~\textsc{ii}] line are on average much weaker in C35 than C32. Recall these fields are at similar Galactocentric distances, but on opposite sides of the Galactic plane, residing near (l, b) = (0.0\td, 1.0\td) and (0.0\td, -1.0\td), respectively. The NGC 6522 and OGLE fields, which are further distant at (0.4\td, -2.1\td) and (1.0\td, -3.8\td), respectively, are dominated by zodiacal emission and quite alike after subtraction. The emission features in the OGLE field are a bit brighter than those of NGC 6522, and its continuum emission beyond 28 \mt rises and slightly diverges from the NGC 6522 spectra. Otherwise, they are quite similar.

\subsection{Variability of the emission features}
\label{sec:refappendix}

Here we discuss the emission features within fields C32 and C35. The tabulated fluxes of all measured quantities are presented in the Appendix. Additionally, maps of all band/line strengths measured within these fields are presented in the Appendix (Figs.~\ref{fig:c32maps},~\ref{fig:c32maps2},~\ref{fig:c35maps} and~\ref{fig:c35maps2}). Due to the limited number of reliable measurements within the OGLE and NGC 6522 fields (c.f. Tables~\ref{table:pahfluxes} and~\ref{table:atomicfluxes}) no such figures are prepared for these sources.

\begin{figure}
    \begin{center}
    \includegraphics[width=1\linewidth] {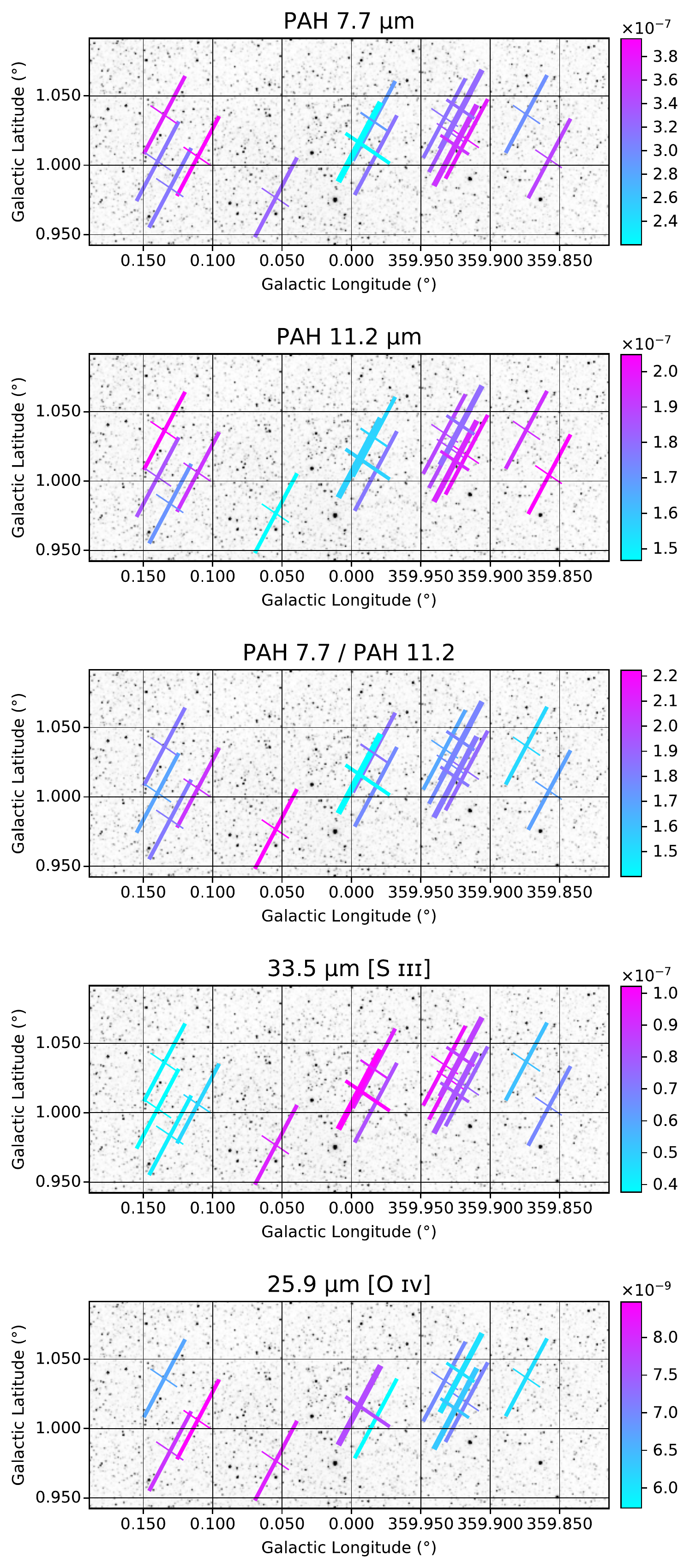}
    \caption{
    Maps of select emission band fluxes in the C32 field, averaged over each aperture and in units of Wm$^{-2}/$sr, overlaid on a 2MASS J-band image. From top to bottom: the 7.7 \mt PAH band, the 11.2 \mt PAH band, the PAH 7.7/11.2 flux ratio, the 33.5 \mt [S~\textsc{iii}] line, and the 25.89 \mt [O~\textsc{iv}] line. The sulphur fine-structure line peaks in the central part of the map, where the PAH emission is generally weakest (roughly coincident with the elevated \ha emission; see Fig.~\ref{fig:irac_c32c35}). Similar figures for the other PAH bands and atomic/molecular lines are found in Figs.~\ref{fig:c32maps} and~\ref{fig:c32maps2}, respectively.
    }
    \label{fig:mapc32}
    \end{center}
\end{figure}

We first examine select emission features in C32 (Fig.~\ref{fig:mapc32}), overlaid on a J-band image from the Two Micron All Sky Survey (2MASS; \citealt{skrutskie2006}). The 7.7 and 11.2 \mt PAH bands are weakest in the central part of the C32 field (near the apertures of C32-9, C32-10 and C32-11) and peak towards the outer regions of the field (near C32-2 and C32-12; c.f., Fig.~\ref{fig:irac_c32c35}). Conversely, the fine-structure lines all peak near the central part of the map, suggesting an anti-correspondence (e.g., the 18.7 \mt [S~\textsc{iii}] emission). Taking into account the three-color image of C32 (Fig.~\ref{fig:rgb_c32}), we infer that there is a correspondence between emission strengths and the \ha channel: where the \ha emission is strong, the atomic lines are brightest and PAH bands weakest. The 25.89 [O~\textsc{iv}] emission is brightest within the \ha emission channel, though it is slightly offset from the 33.5 \mt [S~\textsc{iii}] emission. The 12.8 \mt [Ne~\textsc{ii}] line is detected throughout C32 (Fig.~\ref{fig:c32maps2}), with significantly variability near the \ha channel: the emission is three times brighter in the \ha channel than in the neighboring leftward positions (positions 12, 14, 15 and 16). The 15.5 \mt [Ne~\textsc{iii}] line is much less variable, and is only detected in the central part of the field (in five positions). The [Ne~\textsc{iii}]/[Ne~\textsc{ii}] flux ratio shows no monotonic trend, though note the [Ne~\textsc{iii}] line is relatively noisy in these spectra. Turning to the 17.0 \mt and 28.2 \mt H$_2$ lines, these are weakest in the \ha channel and strongest towards the far-right pointing (C32-1), furthest from the channel.

We prepare a similar figure for C35 (Fig.~\ref{fig:mapc35}), with the full set of maps located in the Appendix (Figs.~\ref{fig:c35maps} and~\ref{fig:c35maps2}). This field has only five positions (two of which, C35-4 and C35-5, share an LL aperture), so our ability to trace smooth variations is limited. What is clear however is that positions 3, 4 and 5, which are coincident with the elevated \ha emission (Fig.~\ref{fig:rgb_c35}), have elevated fine-structure line emission relative to positions 1 and 2. The PAH emission is relatively flat across the field, though perhaps slightly higher in the \ha channel (positions 3, 4 and 5). These positions also exhibit a $\sim$10\% higher PAH 7.7/11.2 \mt flux ratio than the other locations. The 17.0 \mt H$_2$ emission is generally greater towards positions 3, 4 and 5, which is opposite behavior from that of C32.

The OGLE and NGC 6522 spectra are generally too noisy for this type of analysis, though the 34.8 \mt [Si~\textsc{ii}] emission is well detected, peaking strongly at position OGLE-7 within this field. Within NGC 6522, the Si emission is essentially flat across the field, with the exception of possibly NGC 6522-1.

\begin{figure}
\begin{center}
   \includegraphics[width=1\linewidth] {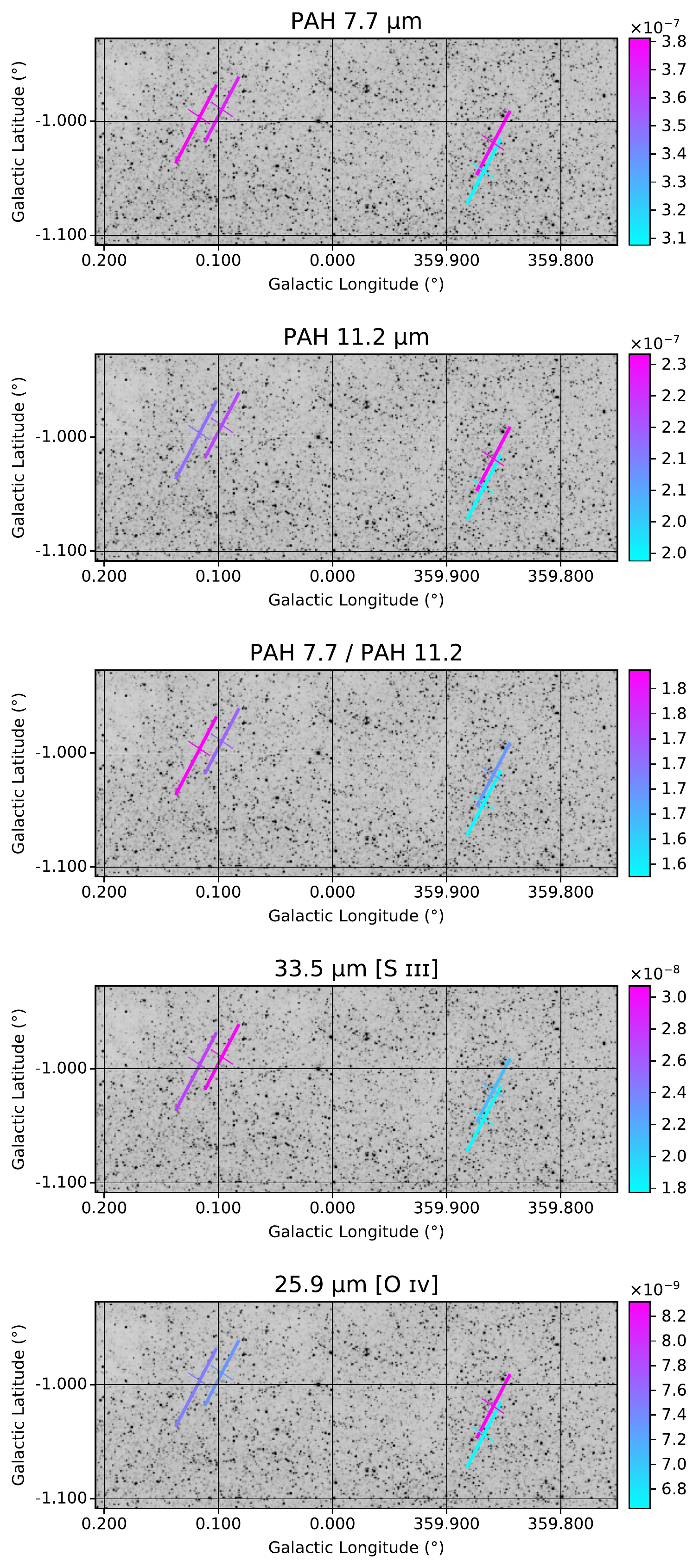}
\caption{Maps of select emission band fluxes in the C35 field, averaged over each aperture and in units of Wm$^{-2}/$sr, overlaid on a 2MASS J-band image. From top to bottom: the 7.7 \mt PAH band, the 11.2 \mt PAH band, the PAH 7.7/11.2 flux ratio, the 33.5 \mt [S~\textsc{iii}] line, and the 25.89 \mt [O~\textsc{iv}] line. The PAH emission varies weakly across the field, while the fine-structure lines are significantly stronger towards the left in this orientation--where elevated \ha emission is present (positions 3, 4 and 5; see Figs.~\ref{fig:irac_c32c35} and~\ref{fig:rgb_c35}). We present similar figures for the other PAH bands and atomic/molecular lines in Figs.~\ref{fig:c35maps} and~\ref{fig:c35maps2}, respectively.}
\label{fig:mapc35}
\end{center}
\end{figure}

\subsection{PAH flux ratio correlations}
\label{sec:corr}

\begin{figure}
	\centering
	\includegraphics[width=1\linewidth]{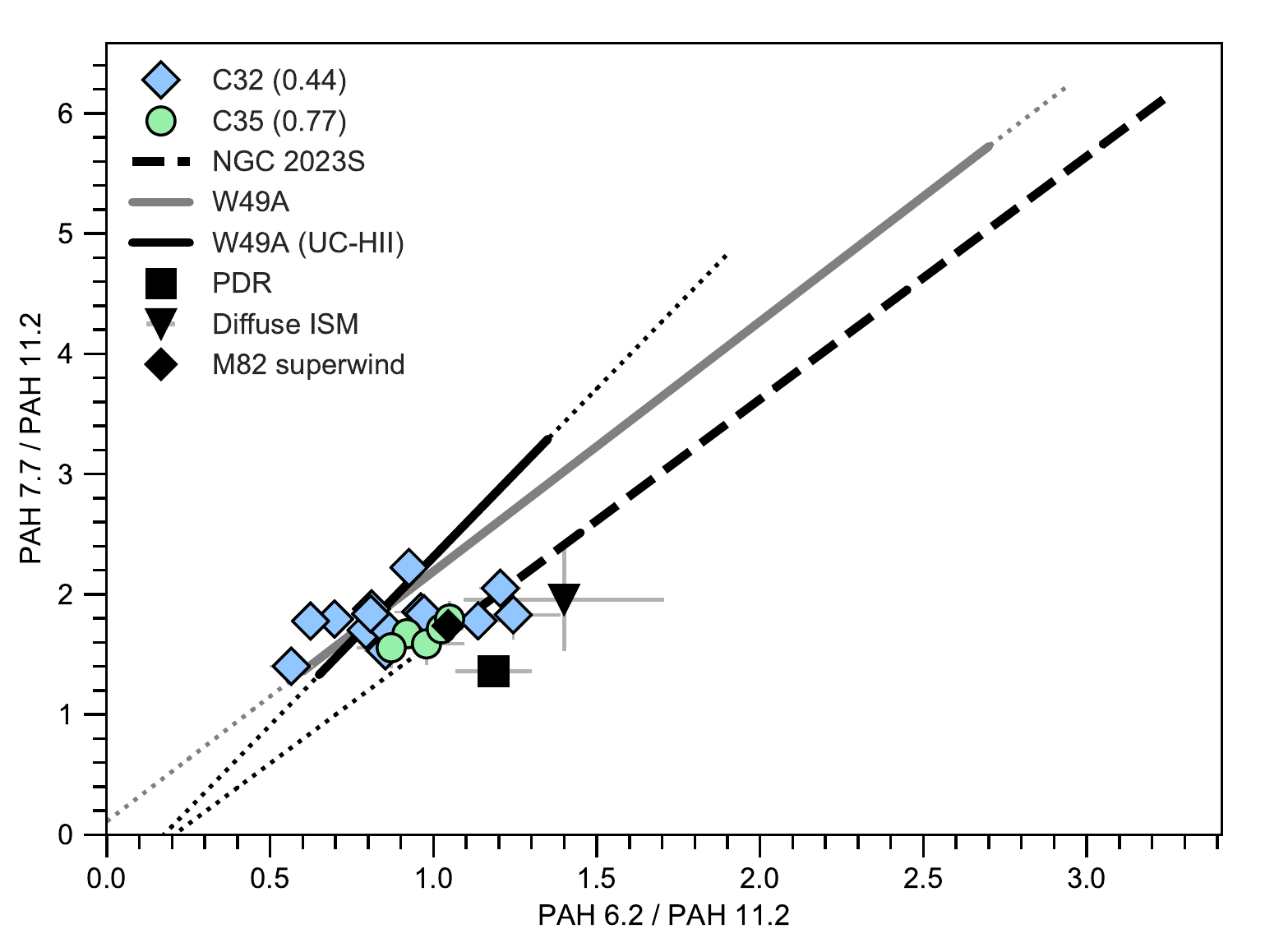}
	\caption{
	Flux correlations between the 6.2 and 7.7 \mt PAH bands, normalized to the 11.2 \mt PAH emission, for C32 (blue diamonds), C35 (green circles) and OGLE fields (red triangles). In parentheses is the weighted Pearson correlation coefficient for each field (omitted if fewer than 3 data points). The lines shown are lines of best fit from \textit{different} data sets: the dashed black line for NGC 2023 \citep{peeters2017}, the solid gray line for W49A \citep{stock2014}, and the solid black line for ultracompact \HII~regions in W49A \citep{stock2014}. The extents of the solid lines indicate the range in the data from which they were originally determined. Extrapolations of these best fits are shown with the light dotted lines. The black symbols correspond to three environments for comparison: the Orion Bar PDR \citep{peeters2002}, the diffuse ISM \citep{bernard1994,boulanger1996a} and a pointing toward the M82 superwind (\citealt{beirao2015}, their region 1). These are explored in Sec.~\ref{sec:enviro_comparison}.
    }
	\label{fig:corr}
\end{figure}

A common method for indirectly tracing systematic variations in PAH populations is by evaluating band PAH flux ratios across environments (e.g., \citealt{galliano2008b}). In Fig.~\ref{fig:corr} we examine the emission strengths of the 6.2 and 7.7 \mt PAH bands using the 11.2 \mt band as a normalization factor. The 6.2 and 7.7 \mt features are strong in ionized PAHs, while the 11.2 \mt band is strong in neutral PAHs. As such, the 6.2/11.2 and 7.7/11.2 ratios trace PAH ionization (e.g., \citealt{allamandola1999}). A weak correlation is observed (with weighted Pearson correlation coefficient $r=0.47$ in C32). Our data span a range in 6.2/11.2 of 0.5-1.2 in C32, 0.9-1.0 in C35, and 0.4-0.6 in OGLE; in contrast, the 7.7/11.2 ratio varies little in our spectra. \citet{peeters2017} analyzed spectral maps of the reflection nebula NGC 2023 and found a high correlation coefficient between the 6.2 and 7.7 \mt bands ($r>0.97$). We plot their line of best fit for comparison in this figure (Fig.~\ref{fig:corr}). Our data exhibit (much) lower 6.2/11.2 flux ratios than in NGC 2023, with the C32 and C35 measurements roughly consistent with an extrapolation of the best fit line towards lower ratios. We also include best fit lines for W49A, a large star-forming region \citep{stock2014}, which is separated into one fit for ultra-compact \HII~regions alone, and one fit for their entire sample (which includes diffuse sight-lines; see their paper for details). Both of their data sets have correlation coefficients $r>0.8$. The W49A 6.2/11.2 flux ratios reach the low ratios that we observe in our sample ($\sim$0.6). The generally weak correlation of the 6.2 and 7.7 \mt bands in our fields toward the Galactic bulge suggests that we are probing relatively small variations in environmental conditions. For instance, the small range in 7.7/11.2 flux ratio we observe should correspond to a relatively narrow range in PAH ionization fraction (see \citealt{galliano2008b}, their Fig.~18). Furthermore we conclude that, relative to NGC 2023 and W49, our environments contain a higher fraction of neutral PAHs. For comparison, we also plot the flux ratios of the Orion Bar PDR \citep{peeters2002}, the diffuse ISM \citep{bernard1994,boulanger1996a} and a pointing toward the superwind of the starburst galaxy M82 (\citealt{beirao2015}, their region 1)\footnote{Note that we remeasured the PAH emission in the M82 spectrum taken from \citet{beirao2015} with a spline continuum for this analysis, as the authors used a different decomposition approach (PAHFIT).}. These sources are generally consistent with our 7.7/11.2 and 6.2/11.2 flux ratios, with the M82 spectrum being the best match to our Galactic bulge flux ratios.

\subsection{Correlations amongst the entire sample}
\label{sec:corrmatrix}

\begin{figure*}
    \centering
    \includegraphics[width=0.9\linewidth]{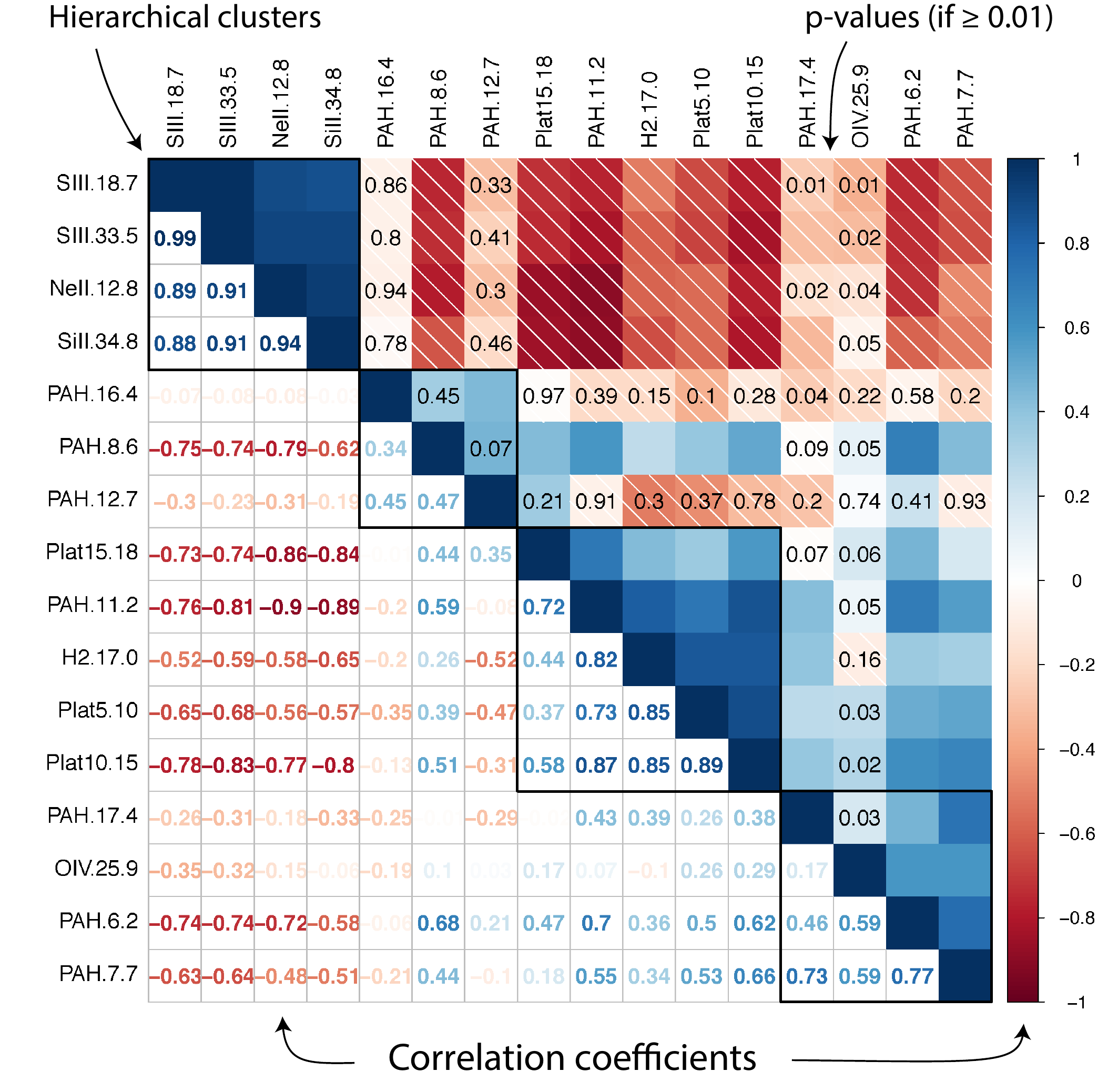}
    \caption{
    Correlation matrix for quantities measured in the spectra of C32 and C35: PAH band fluxes, atomic and molecular line emission, and plateau strengths (see Section~\ref{sec:corrmatrix} for details). The matrix is symmetric, with the Pearson R correlation coefficient presented in the lower half. The upper half is a representation of the correlation coefficient with color-coded squares: positive correlations are blue, negative correlations are red (with diagonal white hatching); the deepness of the color is proportional to the absolute value of the correlation coefficient. P-values are also displayed in the upper half (if p-value $\ge 0.01$), where correlations with p $\le 0.05$ are generally considered significant. The quantities are ordered based on a hierarchical clustering algorithm (black squares). Note the abbreviation ``Plat'' refers to the PAH plateaus.
    }
    \label{fig:corrmatrix}
\end{figure*}

Expanding on the 6.2 and 7.7 \mt band correlation explored in Sec.~\ref{sec:corr}, we now consider correlations between all measured quantities in C32 and C35: PAH bands, plateaus and atomic and molecular emission lines. For this analysis we exclude features with few detections (i.e., the 17.8 \mt PAH band and the 15.5 \mt [Ne~\textsc{iii}] and 28.2 \mt emission lines). We present a correlation matrix summarizing our results in Fig.~\ref{fig:corrmatrix}, ordered by complete-linkage hierarchical clustering. This clustering method computes the maximum (Euclidean) distance between two data points that belong to two different clusters. Based on the matrix, we make a few remarks.

Generally, most PAH features correlate with each other. They also correlate with the plateau emission and the 17.0 \mt H$_2$ line. As an exception, the 12.7 and 16.4 \mt PAH emission bands exhibit a weak correlation ($r=0.45$) with each other but little else. The 12.7 \mt band must be isolated from its blended neighbor at 12.8 \mt ([Ne~\textsc{ii}]), which may explain why there are few statistically significant detections. The solitary nature of the 16.4 \mt band is peculiar, though it may be due to systematic effects in the continuum determination near this location (which is on the rising red wing of the 15-18 \mt plateau).

The fine-structure lines of 12.8 \mt [Ne~\textsc{ii}], 18.7 \mt [S~\textsc{iii}], 33.5 \mt [S~\textsc{iii}] and 34.8 \mt [Si~\textsc{ii}] are highly correlated with each other and anticorrelated with all other quantities. These fine-structure line all originate in ionized gas. The S and Ne ions have comparable ionization potentials (21-23 eV), whereas the Si ion has a potential of approximately 8 eV. The 25.89 \mt [O~\textsc{iv}] ion, which also has an ionization potential of approximately 55 eV, correlates with the 6.2 and 7.7 \mt PAH bands, suggesting it is prominent in environments that favor ionized PAHs.

\subsection{Spectral energy distributions}

Using our photometric images, we have constructed spectral energy distributions (SEDs) for each position in our fields. This was accomplished by sampling the surface brightness of the photometric observations at each pixel in our \textit{Spitzer}/IRS apertures in the following way: if the IRS pixels were entirely contained within a larger pixel of the photometric images, the corresponding surface brightness was adopted; if any IRS pixel overlapped multiple photometric pixels, the mean surface brightness was adopted. In no cases were the IRS pixels larger than the spatial resolution of the photometric images.

\begin{figure}
    \centering
    \subfigure{
    \includegraphics[width=\linewidth] {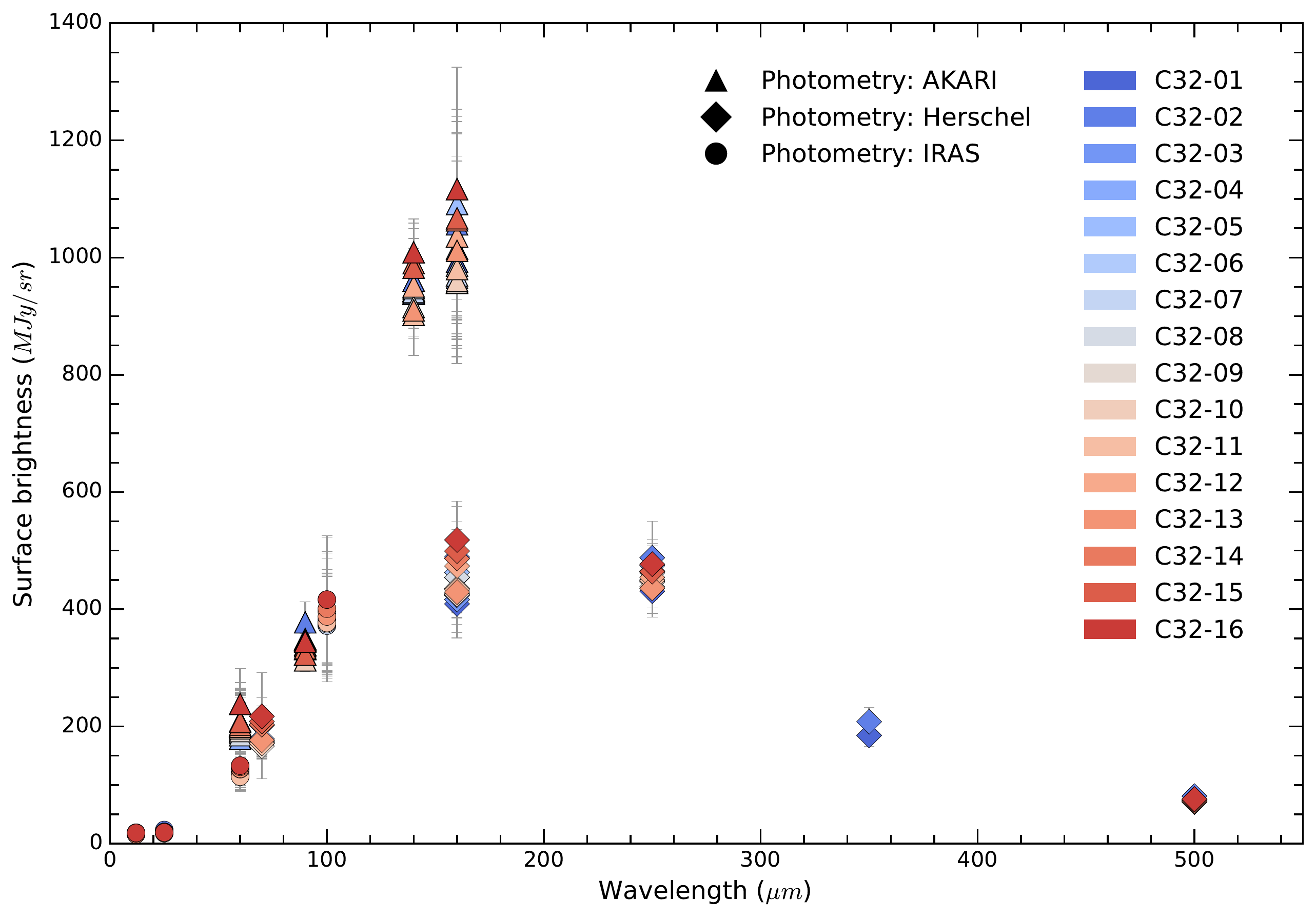}
    \label{fig:sed_c32}
    } \\
    \subfigure{
    \includegraphics[width=\linewidth] {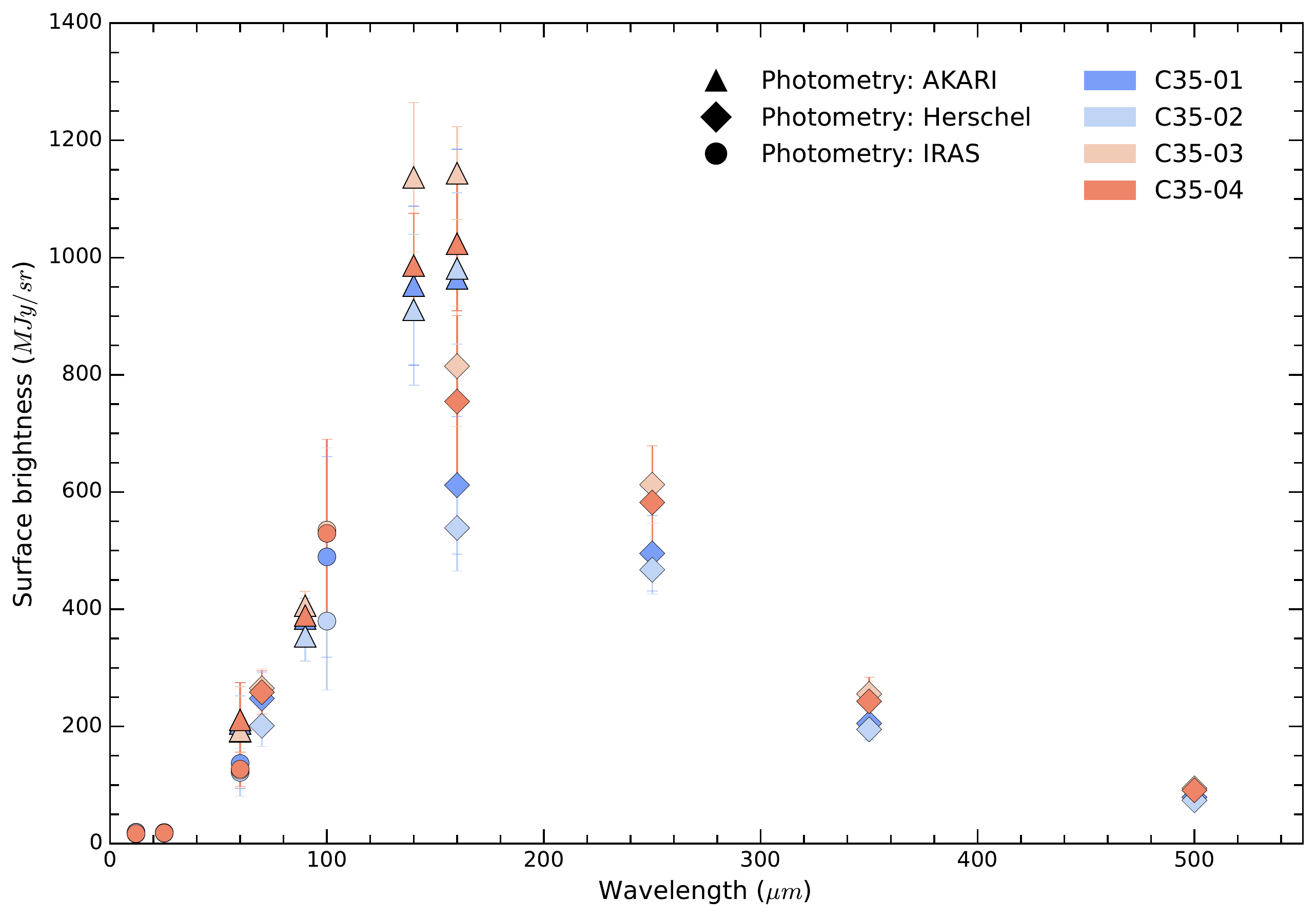}
    \label{fig:sed_c35}
    }
    \caption{
    Spectral energy distributions for C32 (top) and C35 (bottom), constructed with photometric observations from \textit{Herschel} Space Observatory, \textit{Infrared Astronomical Satellite} and \textit{AKARI} (see Table~\ref{table:photometry}). A clear separation between individual positions is present at 160 \m, 250 \mt and 350 \m, particularly in C35.
    }
    \label{fig:seds}
\end{figure}

Fig.~\ref{fig:seds} displays the resulting SEDs for the 16 pointings toward C32 and the four pointings toward C35 (ignoring the fifth C35 position, as it shares the same LL aperture as C35-4). We do not construct SEDs for OGLE and NGC 6522, as there is no coverage from \textit{Herschel} Hi-GAL at these positions (and thus no $>200$ \mt photometry).

Most of the C32 positions have consistent photometric brightnesses within their uncertainties (particularly with their neighboring positions, e.g. \textit{AKARI} 60 \mt and \textit{Herschel} 70 \m), with the stark exception of the measurements near 160 \m. Specifically, the \textit{AKARI} photometry at 140 and 160 \mt are (approximately) twice the surface brightness of the \textit{Herschel}/PACS 160 \mt measurement. The origin of the discrepancy is likely due to the greater calibration uncertainty on the \textit{AKARI} measurements; some residual striping was also observed in the \textit{AKARI} photometric images, which may play a role. The \textit{IRAS}, \textit{AKARI} and \textit{Herschel} measurements all agree at shorter wavelengths ($< 100$ \m). For the purpose of this analysis, we addressed the discrepancy in two ways: first, by including \textit{AKARI}/FIS measurements at 140 and 160 \m; and second, by only including the \textit{Herschel}/PACS measurement at 160 \m. All other measurements are uncontroversial and therefore included to construct the resulting SED.

There is little variation in the C32 photometric measurements, Fig.~\ref{fig:seds}, across the field at both short wavelengths ($<60$ \m) and long wavelengths (350, 500 \m; note however that the coverage of the 350 \mt image is limited, and thus the 350 \mt emission could not be measured at all positions). In the $60-250$ \mt range, surface brightnesses vary between individual C32 positions, sometimes by as much as 50\%. At 160 \m, there appears to be a general, though not strictly monotonic, segregation in brightness of C32 positions west of the channel (toward higher brightnesses) and positions east of the channel (toward lower brightnesses). The \textit{Herschel} and \textit{AKARI} measurements (aside from the 160 \mt filter) show that the C32-1 position is consistently the brightest in the field. This is also true for the \textit{IRAS} measurements, though there is less variation between positions in these data (possibly due to its larger spatial resolution relative to \textit{AKARI} and \textit{Herschel}).

The field of C35 exhibits similar trends: we observe consistent measurements between the three observatories, apart from the $\sim$160 \mt discrepancy. In addition, surface brightness variations within the C35 field are present in the $60-250$ \mt range. Perhaps the most characteristic is that positions C35-1 and C35-2 cluster together in surface brightness, while positions C35-3 and C35-4 are clustered and offset to higher overall brightnesses. Positions 3 and 4 are coincident with elevated \ha emission (Fig.~\ref{fig:irac_c32c35}), which suggests that environmental conditions vary across the C35 field. This separation and clustering can also be seen in several of the C35 maps (Figs.~\ref{fig:c35maps} and~\ref{fig:c35maps2}).

\subsection{Summary}

The spectra of C32 and C35 appear exceptionally similar in continuum shape and PAH feature strength, even after removing the contribution from zodiacal dust. However, spectral variations within the fields are present when examined in detail. We use a composite image of C32 (Fig.~\ref{fig:rgb_c32}) to identify an elevated \ha emission region (or channel). This \ha channel is linked to weak PAH emission and strong fine-structure emission. The 25.89 \mt [O~\textsc{iv}] line, which is a tracer of shocked gas, is detected across the field but is strongest within and near the \ha channel. SEDs show that there are significant surface brightness variations across the field, sometimes by as much as 50\%, indicating variable dust properties. At 160 \mt there appears to be a general trend toward positions east of the channel being brighter than positions west of the channel.

We examine fewer positions within C35 than C32 but find similar trends: the median spectra in C35 are alike but the composite image identifies significant \ha emission near positions 3, 4 and 5 (Fig.~\ref{fig:rgb_c35}). The fine-structure lines peak in this region, as in C32. The 25.89 \mt [O~\textsc{iv}] line is also observed across the field, but no clear variation is apparent between on-channel and off-channel regions, in contrast to the field of C32. Our SEDs show that the C35 positions coincident with the \ha channel are systematically brighter than the other locations (1, 2). The PAH emission strength in C35 does not vary to the degree observed in C32, but there is a correspondence with \ha structure: the 7.7/11.2 PAH ratio is $\sim$10\% higher here when compared to off-channel positions.

A relatively weak correlation is identified between the 6.2 and 7.7 \mt PAH band fluxes across our sample, which is normally one of the strongest PAH correlations measured. This discrepancy likely arises because we are probing a limited range of PAH flux ratios. It is remarkable though that our ratios are similar to those in ultra-compact \HII~regions \citep{stock2014} and do not coincide with those of reflection nebulae and the more diffuse ISM. In this respect, we note that the fine-structure line emission we observe is not similar to those observed toward these ultra-compact \HII~regions with similar PAH ratios: ultra-compact \HII~regions exhibit higher degrees of ionization, with strong 10.5 \mt [S~\textsc{iv}] emission and much stronger 12.8 \mt [Ne~\textsc{ii}] emission than we see in our spectra \citep{stock2014}.

These lines of sight are complex, in general, but we have several tools to help us understand the processes at play: a prominent \ha emission structure, 25.89 \mt [O~\textsc{iv}] emission possibly tracing shocked gas, variable dust grain surface brightnesses and PAH emission variations that show a morphological link to the \ha channel. To understand what environments we are observing along these sight-lines we need to discern whether or not these emission features are simply coincident or linked to the same physical conditions/environments.

\section{Discussion}
\label{sec:discussion}

\subsection{The Galactic Bulge environment}

\begin{figure}
  \centering
    \includegraphics[width=1\linewidth]{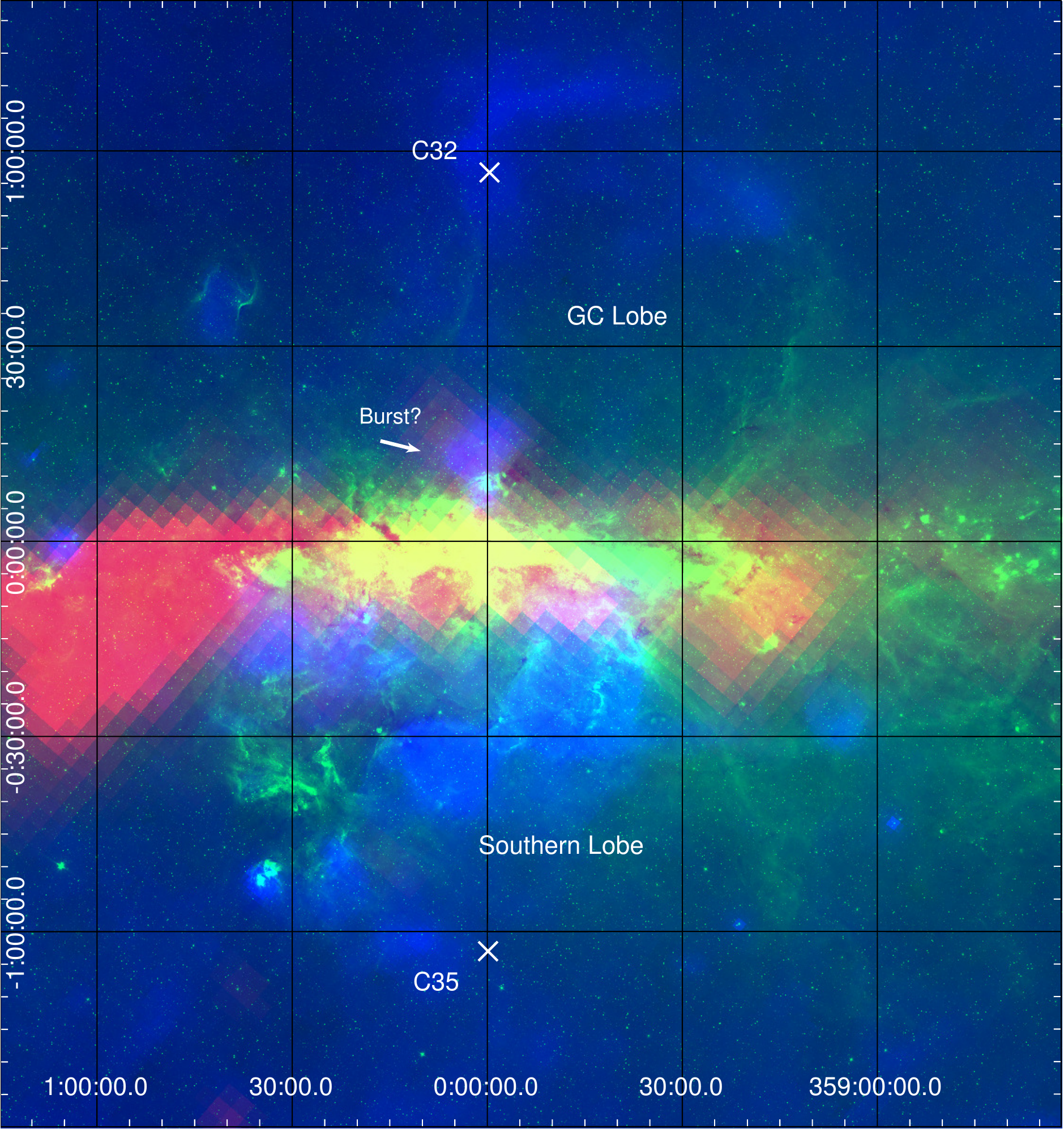}
  \caption{
	A composite image of the Galactic center. The vertical and horizontal axes are Galactic latitude and longitude, respectively. The image consists of \textit{Planck} 857 GHz radio emission (red), 8 \mt \textit{Spitzer}/IRAC PAH emission (green) and SHASSA \ha emission (blue). The C32 and C35 fields are centered on the white crosses (c.f. Figs.~\ref{fig:rgb_c32},~\ref{fig:rgb_c35} for specific aperture locations). The northern loop, as clearly seen in blue, is the Galactic center lobe (GCL); a corresponding southern lobe is also present but less distinctive. Together, the lobes are tilted slightly with respect to the Galactic plane.
    }
  \label{fig:radio}
\end{figure}

We wish to identify which environments we are probing on these sight-lines and the corresponding influence each has on the observed spectra. To address this, we first examine the general Galactic bulge environment. The Galactic plane and fields C32 and C35 are displayed in Fig.~\ref{fig:radio} (spanning approximately $-1.3<b<1.3$). This is a composite image composed of 857 GHz emission measured with the High Frequency Instrument \citep{lamarre2010} of the ESA \textit{Planck} mission \citep{tauber2010}, 8 \mt PAH emission measured with \textit{Spitzer}/IRAC and \ha emission from SHASSA.

North of the plane is the prominent $\Omega$-shaped emission feature known as the Galactic center lobe (GCL), which here can be seen as bound by the \ha and 8 \mt emission arcs. This is a feature approximately $\sim$200~pc in diameter spanning $l=359.2\deg - 0.2\deg$ and $b=0.2\deg - 1.2\deg$, first identified by \citet{sofue1984} from 10 GHz radio continuum emission. It is thought to generally have the shape of a telescope dome \citep{bland2003}, with nested shells of radio line emission, radio continuum emission and dust/PAH emission, respectively, from interior to exterior \citep{law2010}. Quantitatively, the general structure (as measured from the center of the GCL) is: radio line emission from a 15-pc thick shell at radius $r=40$ pc, surrounded by radio continuum emission from a 15-pc thick shell at radius $r=55$ pc, and finally a 5-pc thick dust/PAH shell at radius $r=65$ pc \citep{law2010}.

Calculations suggest the GCL has a mass of 5$\times$10$^6$ M$_\odot$ and an energy content of approximately 10$^{54}$ to 10$^{55}$ ergs \citep{bland2003}. C32 is on the eastern spur of the GCL, towards its highest-latitude boundary. The \ha emission in this region is consistent with emission expectations from radio recombination line studies of local ionized gas \citep{law2009,law2010}.

A complementary southern lobe to the GCL is visible in Fig.~\ref{fig:radio}, previously identified via radio line emission by \citet{alves2015}. While the (northern) GCL exhibits $\sim$0 km s$^{-1}$ velocities, the southern lobe has velocities of $\sim$15 km s$^{-1}$. The entire bipolar structure appears to be inclined roughly 20\td~W of N \citep{bland2003}. In general, the symmetry of the northern and southern lobes suggests a common origin of formation, though in the south the eastern boundary appears to be a relatively complex environment. C35 is on the lobe boundary in the south, similar to the complementary location of C32 in the north. However, due to the overall $\sim$20\td~tilt of the structure, physical conditions towards C32 and C35 may vary. This is reflected in the weaker fine-structure lines observed toward C35 (relative to C32), though the dust continuum and PAH emission strengths seem unwavering between the two fields.

The GCL is generally attributed to an outflow emanating from the Galactic plane roughly 7-10 Myr ago \citep{lutz1999,bland2003}. The starburst model of stellar winds and/or supernovae \citep{veilleux2005} is consistent with the energy requirements implied for the GCL and this is thought to be the most likely formation scenario \citep{law2010}. Calculations indicate the observed dust temperatures on the GCL boundary cannot be sustained by radiative heating alone, suggesting shock and/or turbulent heating may be present at its boundary \citep{bland2003}. The 25.89 \mt [O~\textsc{iv}] line we observe in both C32 \textit{and} C35 spectra is an indication of ongoing shock processing in each environment.

Also present in Fig.~\ref{fig:radio} is a feature in \ha~emission and 867 GHz radio between the Galactic origin and l, b = (0.06, 0.25), approximately---i.e., directed northward out of the plane. Wisps of 8 \mt or \ha emission may be present that provide weak filamentary connections between this feature and the \ha arc near C32, though it is unclear if they are causally linked---it is possible that the feature is simply a foreground object or a burst of some type. On the southern side of the plane, we do not identify a similarly isolated feature. Instead, a large, complex \ha zone (of size $\sim 0.6\deg \times 0.6\deg$) is centered near l, b = (359.9\td, -0.4\td).

\subsection{Comparison by object type}
\label{sec:enviro_comparison}

\begin{figure*}
  \centering
    \includegraphics[width=1\linewidth]{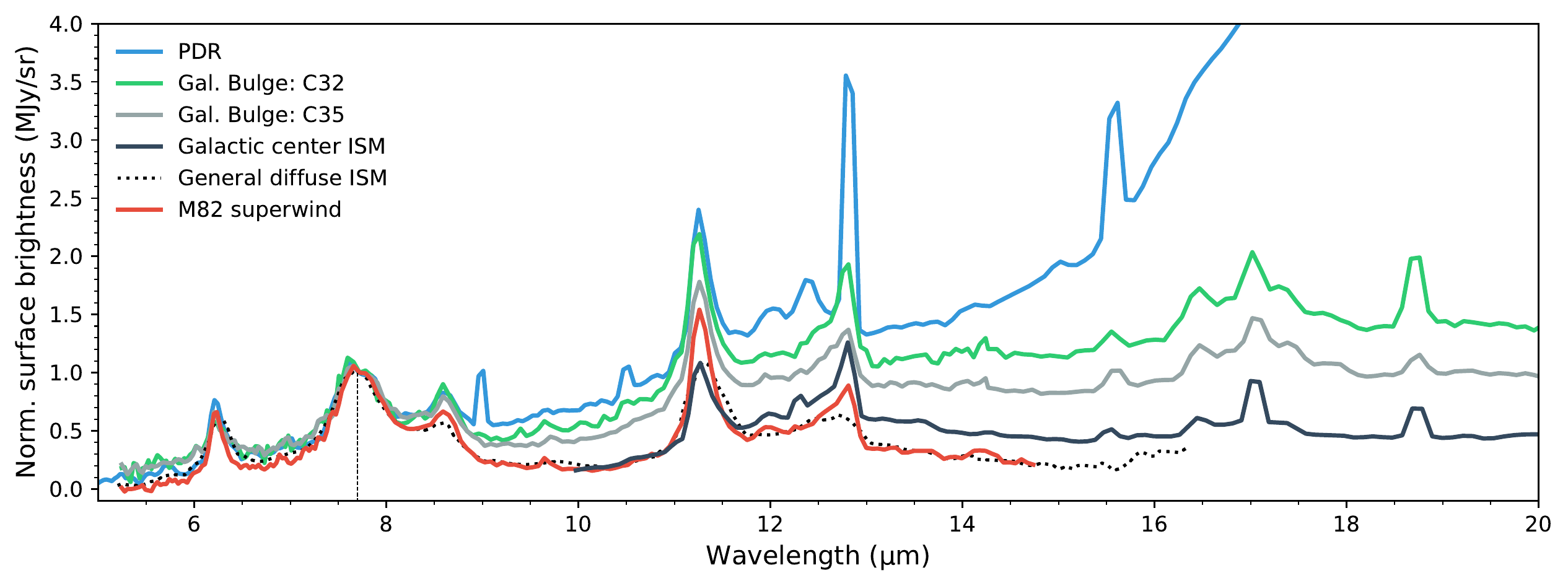}
  \caption{
	A comparison of the mid-IR emission of the C32 and C35 fields (green and grey lines, respectively) and other environments: the Orion Bar PDR (blue), the ISM near the Galactic center (solid black), the diffuse ISM (dotted black) and the superwind of M82 (red). The spectra are normalized to the surface brightness at 7.7 \m. Note that, for clarity, the peaks of the 12.8 \mt [Ne~\textsc{ii}], 17.0 \mt H$_2$ and 18.7 \mt [S~\textsc{iii}] lines have been truncated for the Orion Bar spectrum and Galactic center ISM spectrum.
    }
  \label{fig:pahcompare}
\end{figure*}

\begin{figure*}
  \centering
    \includegraphics[width=0.85\linewidth]{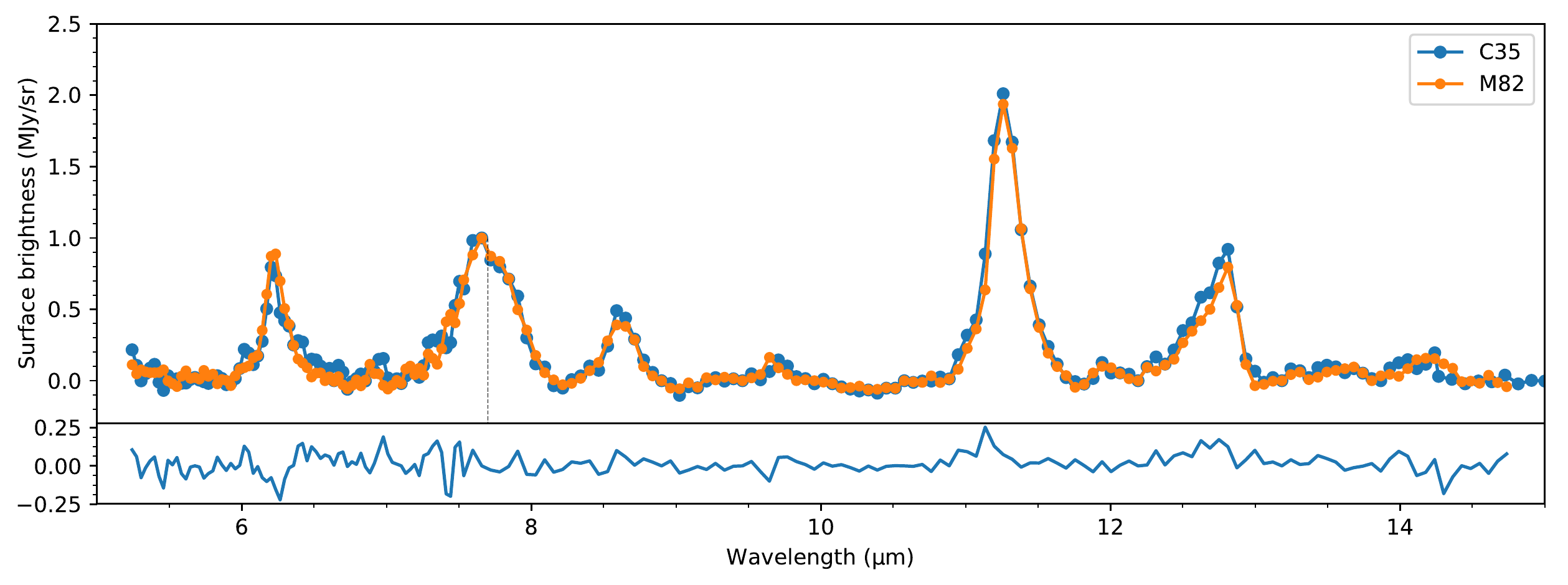}
    \includegraphics[width=0.85\linewidth]{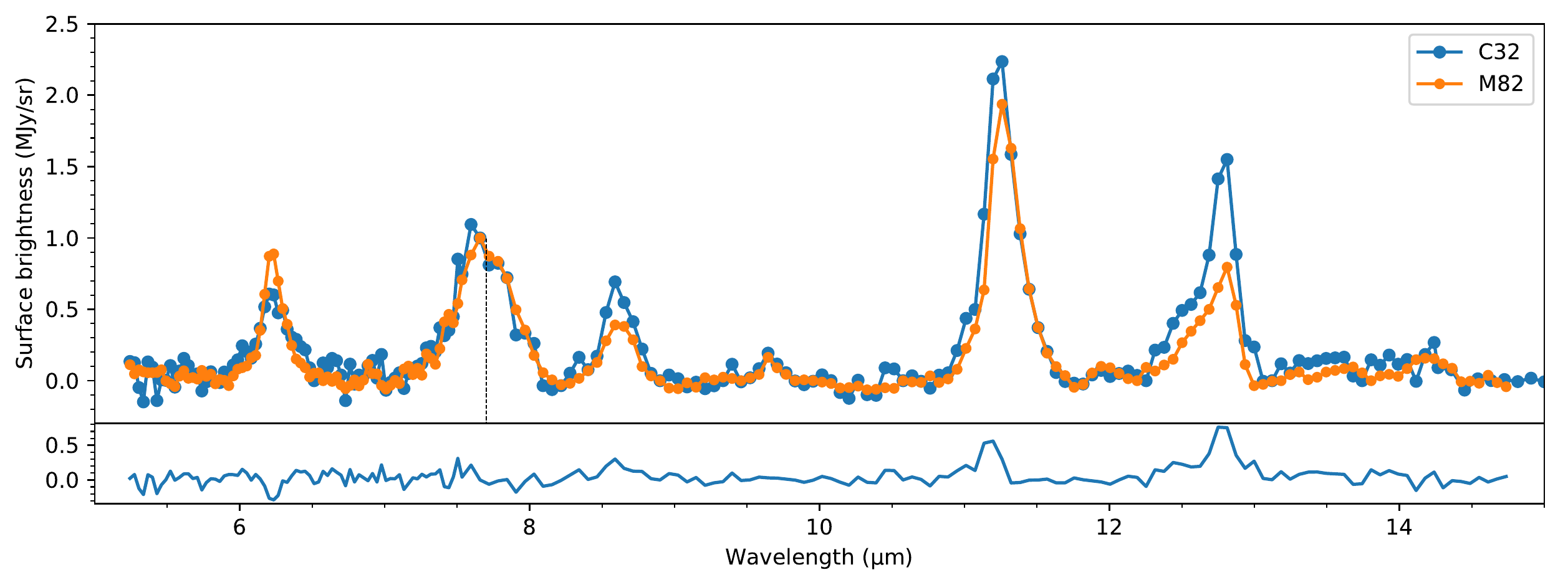}
  \caption{
    A comparison of the M82 spectrum to our C32 and C35 spectra after subtracting a local spline continuum (c.f. Fig.~\ref{fig:pahcompare}). The spectra are normalized to the surface brightness at 7.7 \mt and the residuals are shown below each panel. See Section~\ref{sec:enviro_comparison} for details.
    }
  \label{fig:m82}
\end{figure*}

\begin{figure}
  \centering
    \includegraphics[width=1\linewidth]{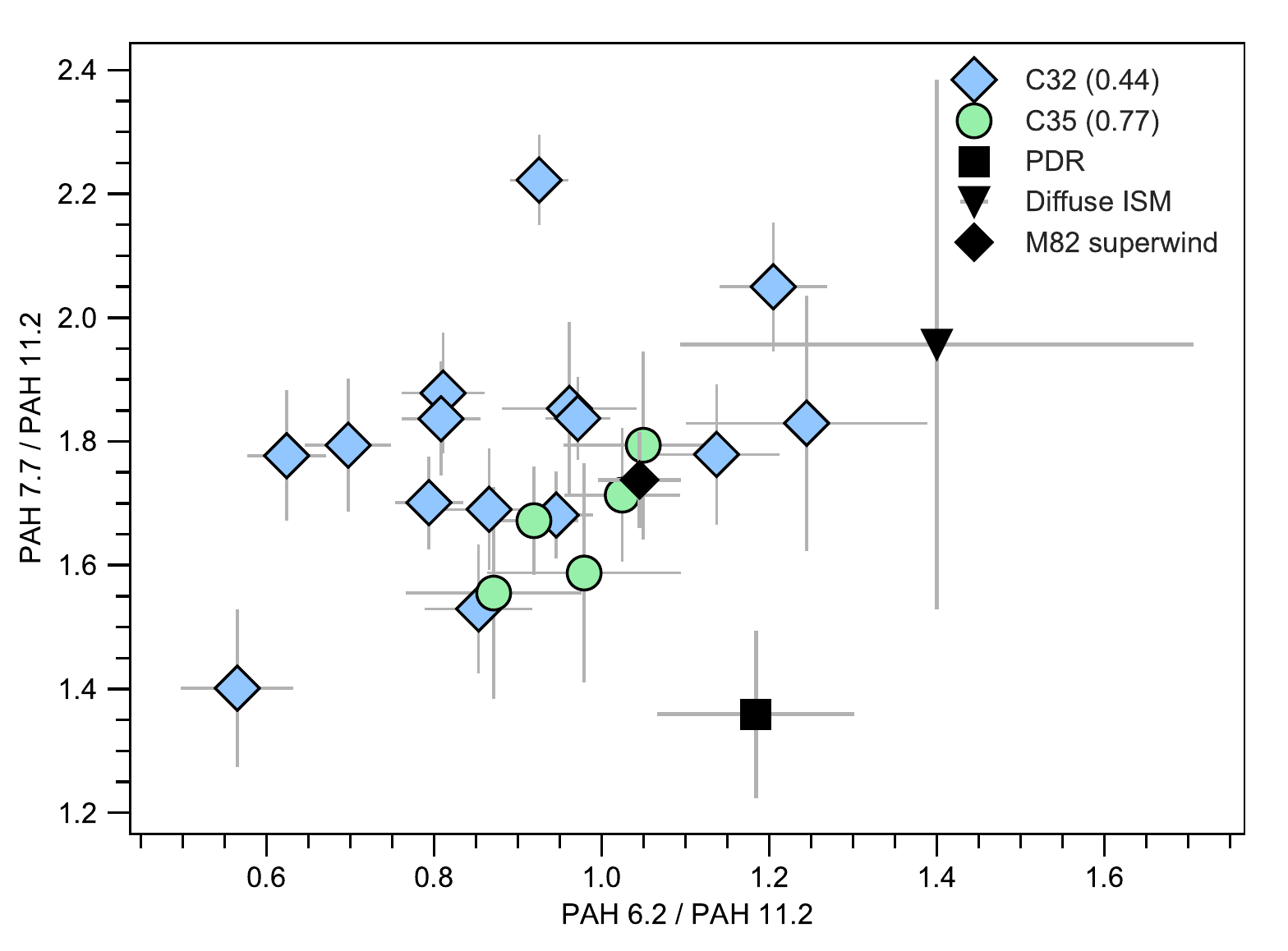}
    \includegraphics[width=1\linewidth]{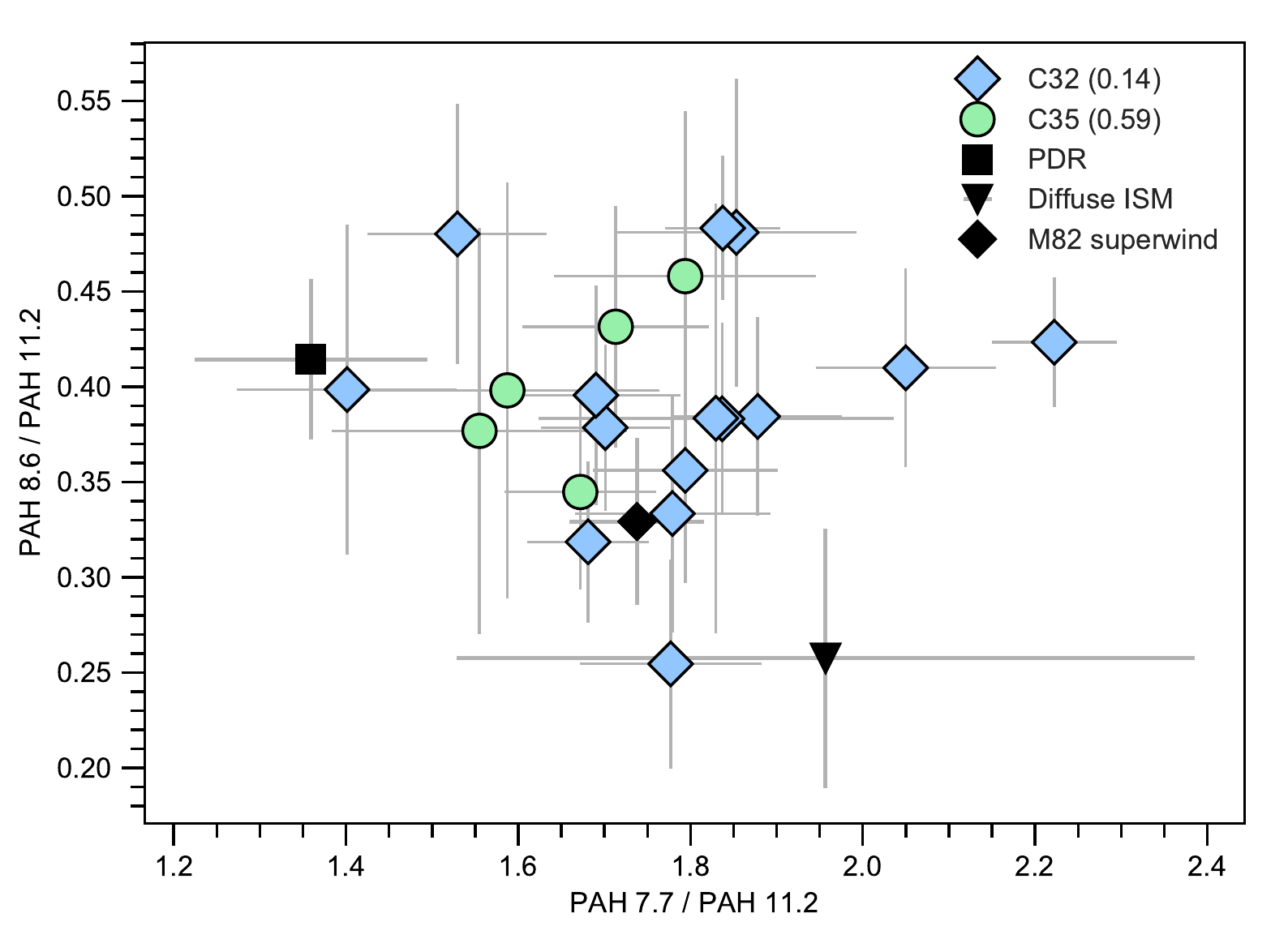}
    \includegraphics[width=1\linewidth]{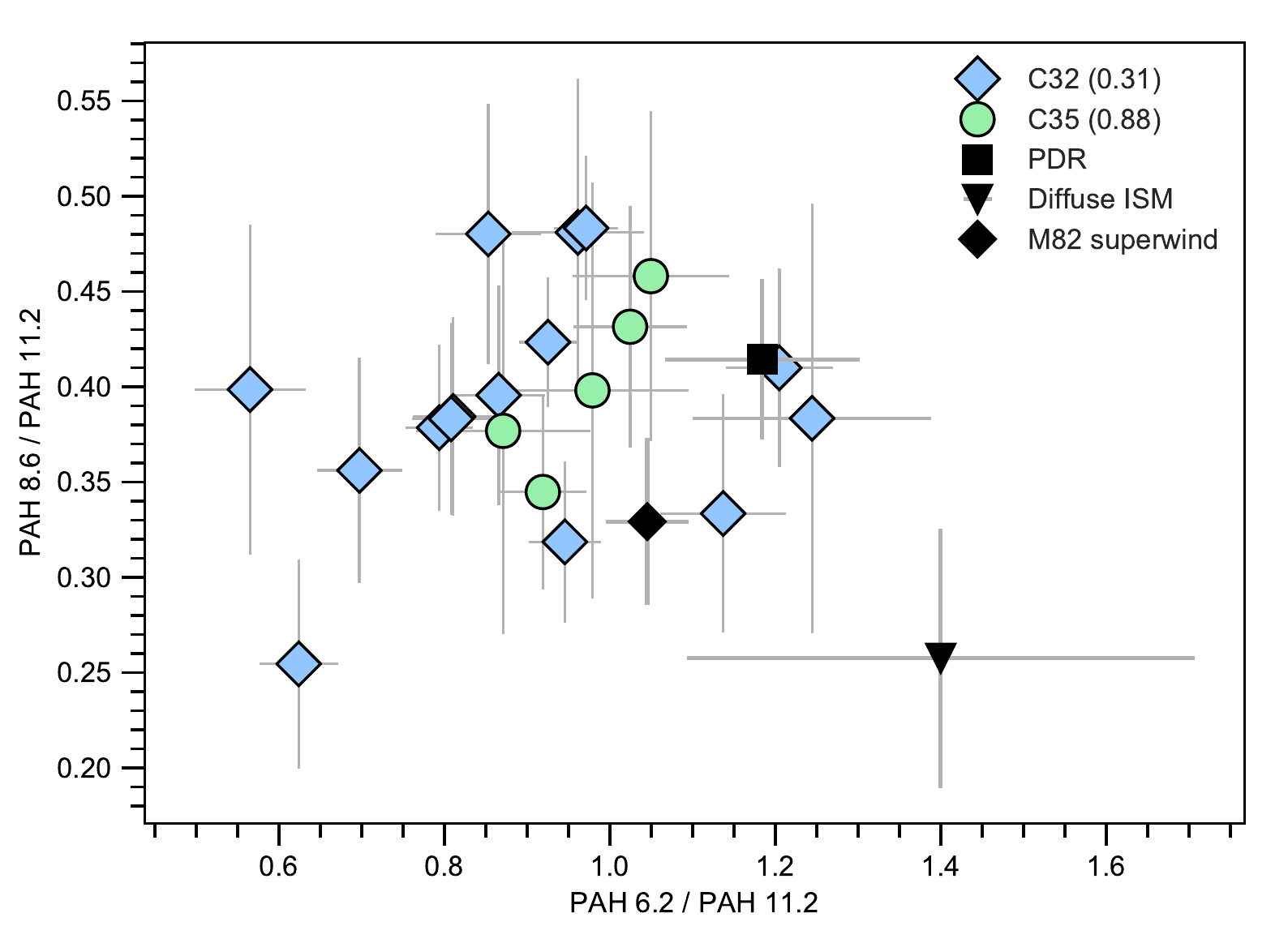}    
  \caption{
	PAH flux ratio correlation plots amongst the 6.2, 7.7 and 8.6 \mt bands for our Galactic bulge sample and three other environments: the Orion Bar (a PDR), diffuse ISM cirrus emission and a pointing toward the M82 superwind. The corresponding mid-IR spectra are presented in Fig.~\ref{fig:pahcompare}. The Pearson correlation coefficient is reported in parentheses for C32 and C35.
    }
  \label{fig:morecorrs}
\end{figure}

To better understand our observations toward the Galactic bulge, we compare our spectra to spectra representative of distinct astrophysical environments in Fig.~\ref{fig:pahcompare}.

\paragraph{Mid-IR.} For the mid-IR comparison, we include i) a spectrum of the Orion Bar, a bright star-forming region \citep{peeters2002}, with data acquired with the Short Wavelength Spectrometer (SWS; \citealt{degraauw1996}) on board the Infrared Space Observatory (ISO; \citealt{kessler1996}); ii) a spectrum of high galactic latitude cirrus clouds to represent the diffuse ISM \citep{bernard1994,boulanger1996a}, with data acquired with ISOCAM \citep{boulanger1996b}; iii) a \textit{Spitzer}/IRS ISM pointing from near the Galactic center located at (l, b) = (0.1152\td, 0.2345\td) (\citealt{simpson2007}, their position 38); and iv) a spectrum of the superwind in the starburst galaxy M82 to represent a galactic wind driven by star formation activity (\citealt{beirao2015}, their region 1) positioned 200\arcsec~from the center of M82 along its minor axis\footnote{While the positions of our FOVs at the distance of M82 (3.3Mpc) correspond to approximately 9-19\arcsec\, along M82's minor axis, the superwind of M82 is likely more extended. For example, \citet{grimes2005} found that the size of the galactic wind in X-ray emission correlates with FIR luminosity while \citet{mccormick2013} reports a roughly constant ratio of the minor axis scale height of the 8 $\mu$m IRAC emission to the FIR luminosity. These relationships indicate that M82's galactic wind is 4-30 times more extended than that of the Milky Way. The corresponding M82 minor axis positions then range from 35\arcsec\, to 570\arcsec.}).

The spectra are normalized to the peak surface brightness at 7.7 \mt for the sake of comparison (Fig.~\ref{fig:pahcompare}). Qualitatively, the spectra of C32, C35 and the Galactic center ISM pointing have similar continuum shapes and PAH features. The diffuse ISM and M82 super wind spectra have a weaker underlying continuum but seemingly comparable PAH emission strengths. The PDR spectrum rises steeply beyond 15 \m, diverging from these spectra, yet exhibits similar PAH band strengths.

Focusing solely on the PAH emission, the spectrum of the M82 galactic wind is a surprisingly close match to the spectrum of C35 (see Fig.~\ref{fig:m82}, upper panel). The relative PAH strengths and profiles are identical. The continuum-subtracted spectrum of C32 is similar to M82 (Fig.~\ref{fig:m82}, lower panel), though the PAH ratios are slightly different -- and the 12.8 \mt [Ne~\textsc{ii}] line is clearly present in the former but not the latter. We therefore conclude that the wind-swept PAHs in these regions are exposed to comparable physical conditions, despite the differences in the underlying continua (Fig.~\ref{fig:pahcompare}). The PAH emission in the region 1 position of the M82 superwind is quite similar to other positions north of the plane (regions 5, 6, 7 and 9 in Fig.~4 of \citealt{beirao2015}). The authors show that the 6.2/7.7 flux ratio is very similar for all of these regions. Strong PAHs are also detected in regions 3, 8, 10, 11 and 12 (the latter three of which are south of the plane) but display a flatter continuum. The authors suggest that the northern wind, of which region 1 is a member, contains PAHs that are larger and more ionized when compared to PAHs in the southern wind or the starburst disc.

To quantify the PAH feature strengths we examine correlations between the 6.2, 7.7 and 8.6 \mt PAH bands in Fig.~\ref{fig:morecorrs}. Each is normalized to the 11.2 \mt feature. The diffuse ISM pointing exhibits a somewhat depressed 8.6/11.2 ratio when compared to our Galactic bulge observations, while Orion Bar has a relatively low 7.7/11.2 ratio when compared with the sample. The M82 superwind position is a close match to our bulge spectra for all three PAH flux ratios. The C35 ratios tend to be closely clustered relative to the span in flux ratios of the C32 field, which may suggest that physical conditions (and correspondingly PAH band fluxes) are changing more rapidly across the C32 field than C35. The OGLE bulge spectra have systematically low 6.2/11.2 and 7.7/11.2 ratios when compared with C32 and C35, which may be a result of the OGLE field being further from the Galactic center---these ratios generally trace PAH ionization, and thus a lower 6.2/11.2 and 7.7/11.2 PAH ratio is consistent with the weak fine-structure lines are observed in the OGLE field (Fig.~\ref{fig:specall4}).

\paragraph{Far-IR.} We implement a modified blackbody fit to our SEDs, $I_{\nu} \propto \nu^{\beta} B_{\nu}(T)$, to characterize the properties of the thermal dust in our fields. $B_{\nu}$ is Planck's function, $T$ the dust temperature and $\beta$ the spectral emissivity index (similar to the approach of \citealt{arab2012}). We fit only the photometric observations past 50 \m, which characterizes the emission of large grains in thermal equilibrium (i.e., we do not account for the emission of stochastically heated PAHs and very small grains). A fit is performed for each pixel in the C32 and C35 fields. Per the photometric discrepancy between \textit{AKARI} and \textit{Herschel}/PACS, we use only the latter for our blackbody fitting.

\begin{figure}
  \centering
    \includegraphics[width=1\linewidth]{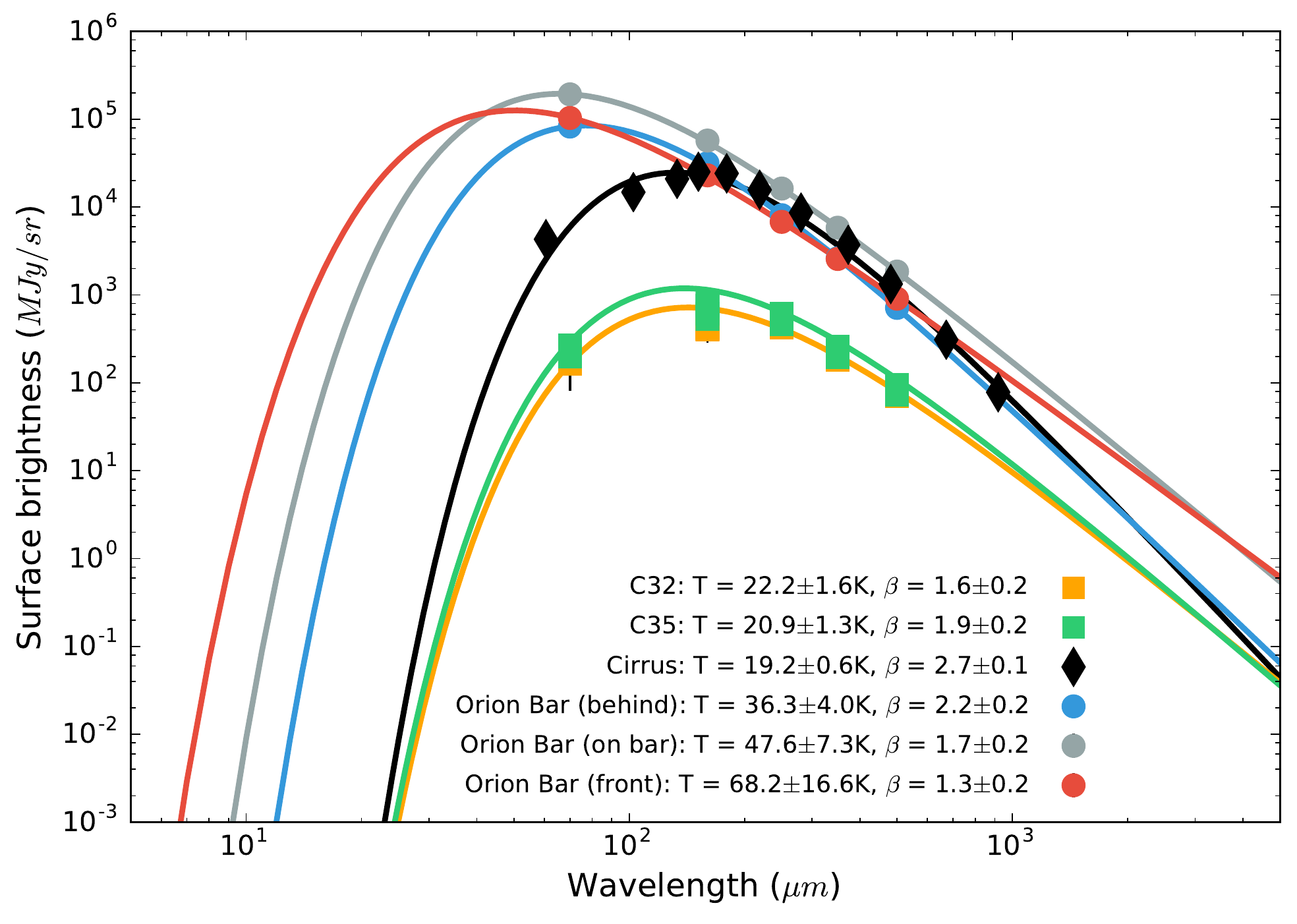}
  \caption{
	Spectral energy distributions for C32, C35, the diffuse ISM (high galactic latitude cirrus) and Orion bar are fit with a modified blackbody. Here, ``front" refers to a position in front of the Orion Bar (i.e., closer to the exciting star), and conversely, ``behind" is further away from the exciting star than the Bar itself. The fit parameters are included in the legend.
    }
  \label{fig:c1}
\end{figure}

The results of this process are presented in Fig.~\ref{fig:c1}. We determine a mean dust temperature of 22.2$\pm$1.6K in C32 and 20.9$\pm$1.3K in C35, which are consistent with each other. The $\beta$ parameter is 1.6$\pm$0.2 in C32 and 1.9$\pm$0.2 in C35, consistent within their uncertainties. We also fit the SED of diffuse cirrus emission from \textit{COBE} satellite observations \citep{bernard1994,boulanger1996a}. The diffuse cirrus emission has a comparable dust temperature to C32 and C35 (19.2$\pm$0.6K), but the fit returns a significantly higher $\beta$ value (2.7$\pm$0.1). We also fit the FIR emission for Orion Bar, known for its edge-on view of the transition from \HII~region to molecular cloud \citep{tielens1993}. We examine SEDs for three positions in Orion Bar previously identified by \citet{arab2012}, their Fig.~4: on the bar itself, 30\arcsec~ahead of the bar (closer to the illuminating source), and 38 \arcsec~further behind the bar (further distant from the illuminating source). The emission in Orion is very bright, dwarfing the C32 and C35 emission by roughly three orders of magnitude at $\sim70$ \m. The Orion FIR continuum monotonically decreases beyond 70 \m, indicating that we miss the peak emission wavelength and thus hotter grains are present in Orion than in the other environments. Our blackbody fits show that grain temperatures in front of, on and behind the bar are approximately 68.2$\pm$16.6K, 47.6$\pm$7.3K and 36.3$\pm$4.0K, respectively. \citet{arab2012} report for the same positions dust temperatures of $70.6\pm10.5$ K, $48.8\pm4.0$ K and $37.1\pm2.5$ K, respectively, which are in good agreement.

\vspace{4mm}
We conclude that the thermal dust we observe toward C32 and C35 is very akin to the dust observed towards high galactic latitude, diffuse cirrus clouds. Moreover, the PAH correlation plots indicate that the YSO and diffuse ISM pointings have similar (though not identical) PAH feature strength ratios to our bulge spectra.

\subsection{PAH size comparison}

\begin{figure}
  \centering
    \includegraphics[width=1\linewidth]{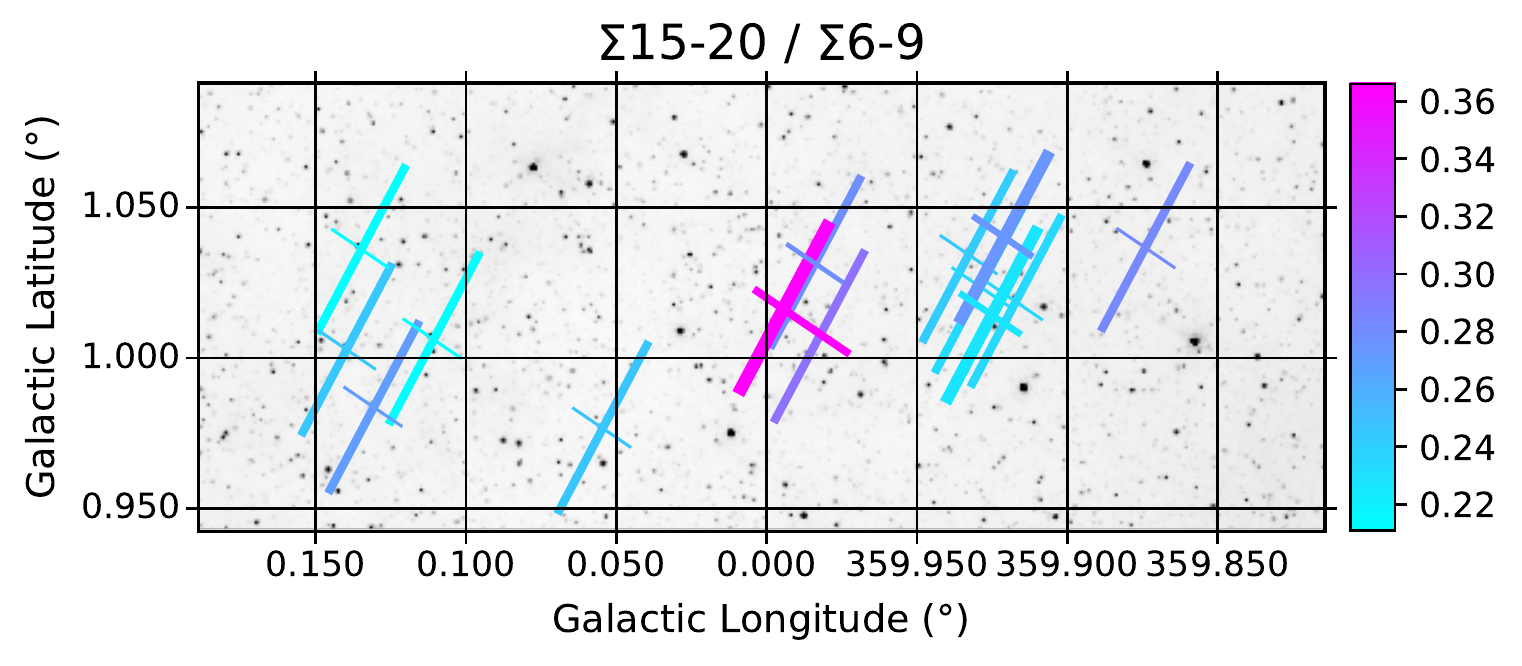}
    \includegraphics[width=1\linewidth]{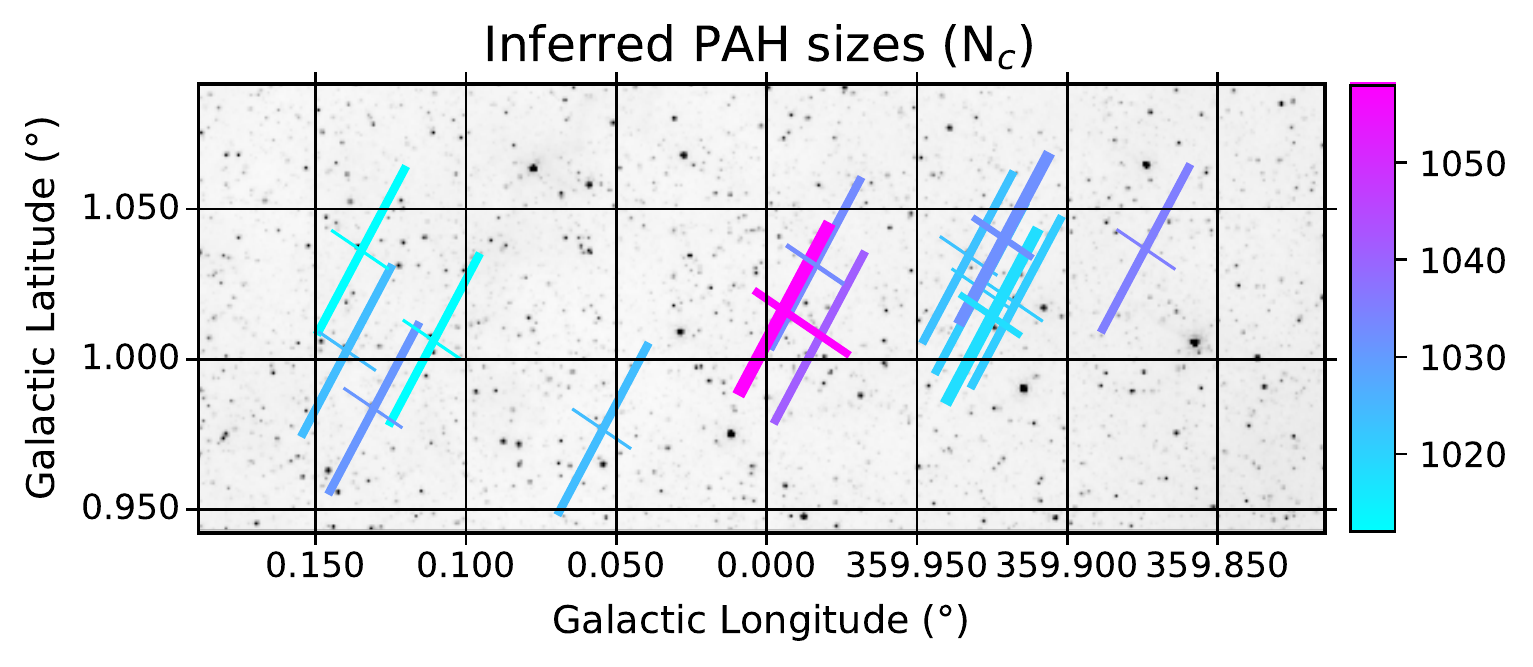}
  \caption{
    Estimating average PAH sizes within the field of C32. \textit{Top:} The ratio of the 15-20 \mt PAH features to the 6-9 \mt features is overlaid on a 2MASS J-band image. Higher values of this ratio imply larger average PAH sizes, as described and quantified by \citet{boersma2010}. \textit{Bottom:} The inferred average PAH sizes in C32. Note that mean sizes are generally higher where there is increased \ha emission, suggesting PAH processing (e.g., the destruction of smaller PAHs).
    }
  \label{fig:pahsizes}
\end{figure}

\citet{boersma2010} determined a relationship between mean PAH size and the ratio of the 15-20 \mt PAH band flux to the 6-9 \mt PAH band flux (their Fig.~19), based on the NASA Ames PAH IR Spectroscopic Database\footnote{\url{www.astrochemistry.org/pahdb/}} \citep[PAHdb;][]{bauschlicher2010,boersma2014_amesdb}. Note that the 15-20 \mt bands reflect C-C-C vibrational modes, while the 6-9 \mt bands are associated with C-C vibrations (see \citealt{boersma2010} for assumptions and methodology). \citet{tappe2012} examined \textit{Spitzer}/IRS spectra of the supernova remnant N132D in the Large Magellanic Cloud. PAH emission is clearly present in multiple positions in and around the blast wave in N132D, but the 15-20 \mt PAH bands are not present on the shock boundary itself, suggesting their carriers are destroyed. By using the approach of \citet{boersma2010}, \citet{tappe2012} examined the PAH size characteristics towards N132D. These authors find mean PAH sizes of N$_c = 4000-6000$, which is larger than is observed in more traditional environments.

We perform the same analysis for C32 and C35. In C32, the 15-20/6-9 ratio varies between approximately $0.20-0.35$ (Fig.~\ref{fig:pahsizes}), comparable to the mean values measured by \citet{boersma2010} and much less than the ratios of $\sim10$ for supernova remnant N132D \citep{tappe2012}. Note that the sample of \citet{boersma2010} contains a mixture of reflection nebulae, \HII~regions, planetary nebula and an average galaxy spectrum prepared by \citet{smith2007} from the Spitzer Infrared Nearby Galaxies Survey (SINGS, \citealt{kennicutt2003}. Our ratios correspond to PAH sizes of approximately N$_c = 1010 - 1060$. The largest mean PAH sizes in C32 are generally coincident with the \ha channel toward the center of the field. Similar 15-20/6-9 ratios are observed in C35 (approximately between $0.21-0.25$, resulting in PAH sizes of roughly N$_c = 1010 - 1025$), but it is mostly homogeneous across the field, with perhaps a smaller mean size in position 3 where there is elevated \ha emission (in contrast to the trend seen in C32).

We also compare our observations to those of supernova remnant N63A \citep{caulet2012}. From the observed positions, we selected the ``NE'' and ``SE'' lobes, based on the similarity of C32 and C35 to these regions in atomic line diagnostic plots (their Fig.~9). On the NE and SE lobes, the 15-20/6-9 ratio is approximately 0.116 and 0.058, respectively. These correspond (very roughly) to PAH sizes of N$_c \sim700$ and $\sim520$, respectively. 

PAHs are also present in the superwind of starburst galaxy M82 \citep{beirao2015}. The implied PAH sizes from the 15-20/6-9 ratio generally are in the N$_c\sim700-1000$ range, with perhaps two or three cresting N$_c=1000$.

In comparison with supernova remnants, we find that the mean PAH sizes in C32 and C35 are much smaller than those observed in supernova remnant N132D \citep{tappe2012} and significantly larger than PAHs in supernova remnant N63A \citep{caulet2012}. The infrared spectra of supernova remnants can be very complex and considering the variation in PAH sizes between remnants N132D and N63A, it's clear that PAH processing is highly dependent on the local conditions and energetics. In contrast, the derived PAH sizes for C32 and C35 are similar to those determined for all but the supernova remnant objects and are thus very typical. On a smaller spatial scale, we note that within the C32 field, larger typical PAH sizes are observed in the central region/\ha channel, which suggests that smaller PAHs are been preferentially destroyed in this region.

\subsection{The 17 \mt PAH plateau}

The emission plateau centered near 17 \mt was originally identified by \cite{vankerckhoven2000} using \textit{ISO} observations of \HII~regions, YSOs and evolved stars. In these data, the plateau emission generally spans 15-20 \m. These authors report that, overall, the shape of the plateau is very similar between sources, with few sources exhibiting discrete emission features on top of the broad plateau, most noticeably at 16.4 \m. Thanks to \textit{Spitzer}'s superb sensitivity, the spectral details of the 15-20 \mt PAH emission features have been further revealed (e.g., \citealt{werner2004,peeters2004b,sellgren2007,boersma2010,peeters2012,shannon2015}). Most commonly observed are a set of discrete emission features at 15.8, 16.4, 17.4 and 17.8 \mt located on top of a broader emission band centered at 17 \m. The latter plateau specifically appears to span 15-18 um and is thus flanked by the 15.8 and 17.8 \mt bands. This is very distinct from the broad, nearly flat-topped plateau from 15-20 \mt observed in \HII regions \citep{vankerckhoven2000,peeters2006}. This dichotomous behavior is somewhat mysterious. Here we examine the nature of the plateau within our bulge observations.

Prior to any continuum subtraction, the spectra in our sample exhibit strong 15-18 \mt bands and plateau emission in the C32 and C35 fields (Fig.~\ref{fig:specall4}). In NGC 6522, the plateau emission appears to extend to approximately 20 \m, with only the slightest hint of 15-18 \mt emission bands. The OGLE field shows a mixture of both the broad plateau and weak 15-18 \mt emission bands.

\begin{figure}
  \centering
    \includegraphics[width=1\linewidth]{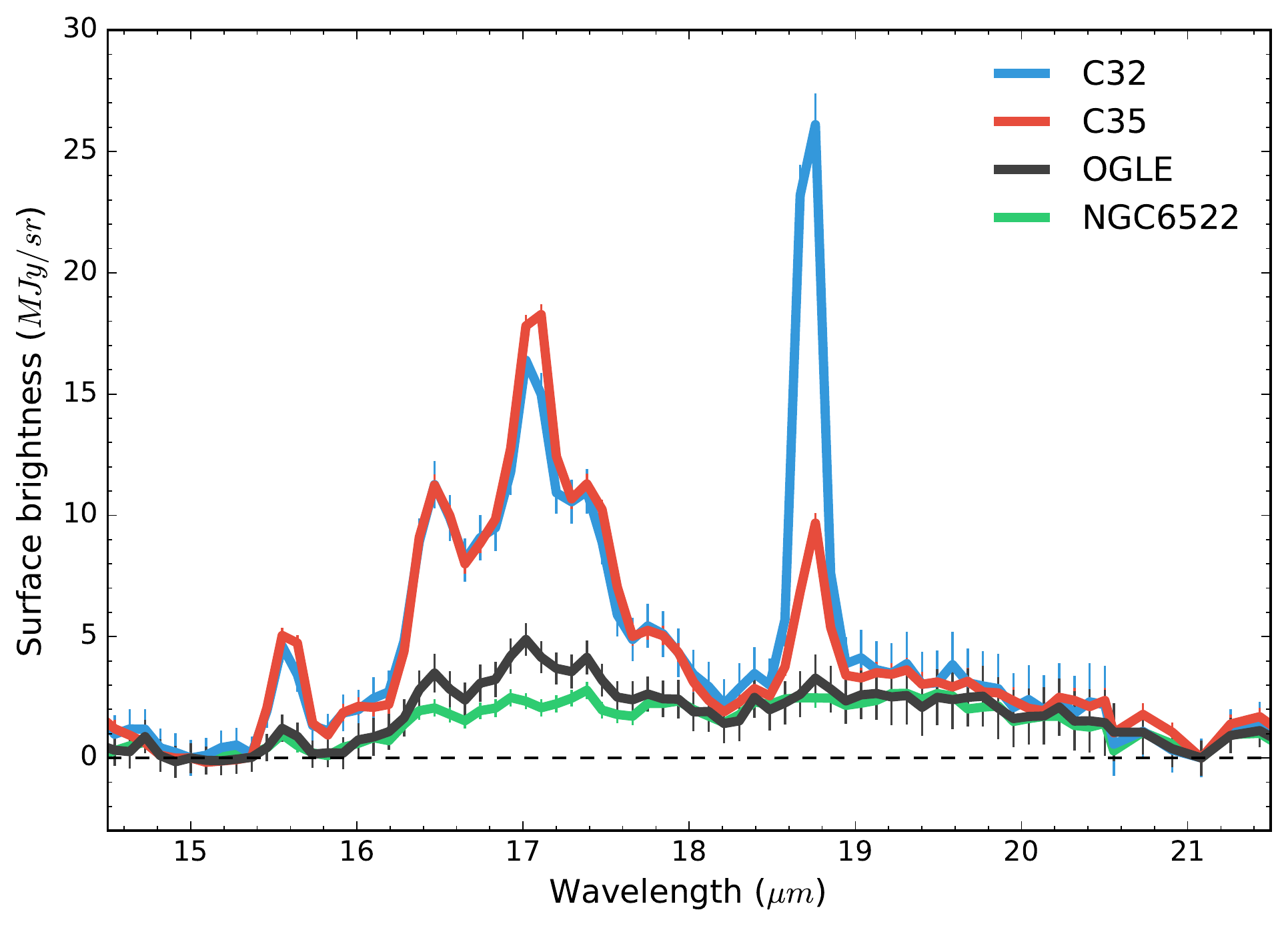}
    \caption{The median emission in the 15-20 \mt range for each field after subtracting a continuum to isolate the plateaus (see Fig.~\ref{fig:spectrum1}). Emission near 17 \mt dominates the 15-20 \mt region of C32 and C35, but it is of comparable surface brightness to the 18-20 \mt emission in the OGLE and NGC 6522 fields.}
  \label{fig:plateaus}
\end{figure}

We examine the continuum-subtracted PAH emission of our observations in Fig.~\ref{fig:plateaus} (see Fig.~\ref{fig:spectrum1} for a sample continuum). Residual emission is present in NGC 6522 between approximately 16 and 20.5 \m, despite the lack of a prominent 17 \mt bump. This resembles the broad 15-20 \mt plateau emission seen towards \HII~regions \citep{vankerckhoven2000,peeters2004b,peeters2006}. The emission in the C32 and C35 fields conversely is dominated by the discrete PAH emission features and the 15-18 \mt plateau centered on 17 \m. The PAH emission in the OGLE field is somewhat intermediate, with a weak but discernible 17 \mt bump. Note however that despite these differences in the 15-18 \mt PAH emission, all four fields have comparable (residual) emission between 18 and 20.5 \mt after continuum subtraction (excluding the 18.7 \mt [S~\textsc{iii}] line). It is possible that C$_{60}$ emission near 18.9 \mt could be present in these sources \citep{cami2010,sellgren2010}, but it likely does not affect the residual emission we observe for two reasons: (1) the 17.4 \mt band is very weak, suggesting a weak 18.9 \mt C$_{60}$ band, if any, and (2) the full-width at half-maximum of the 18.9 \mt C$_{60}$ feature is not nearly as wide as the residual/plateau emission. In summary, this suggests that in addition to the visible 15-18 \mt plateau emission (as clearly seen in the C32 and C35 fields) a second broad emission component may be present. It is unclear if this second component spans the entire 16-20.5 \mt range, and thus lies underneath the 15-18 \mt plateau emission, or whether it is only adjacent, spanning 18-20.5 \m.

The origin of the dual nature of the plateau emission (i.e., extending to 18 \mt vs. 20 \m) is unknown. One possibility is that it represents a systematic effect in these analyses: consider that the plateau shape and extent is influenced by the way in which the underlying continuum is drawn. However, in this work it appears that the 15-18 \mt plateau (as seen in C32 and C35) can be observed prior to continuum subtraction, as can the 15-20 \mt plateau (as seen in NGC 6522), which minimizes this bias to some degree. Another possible systematic effect is that the Spitzer/IRS modules LL1 and LL2 overlap in the 20-21 \mt range, so a calibration offset may affect how the continuum is defined and thus the plateau shape. However, the \textit{ISO} spectra had no such stitching point near this spectral range so it is perhaps an unlikely source of the discrepancy.

Density functional theory calculations of C-C-C PAH bending modes (which are the vibrational modes attributed to this emission region) predict that the 15-20 \mt region is dominated in emission intensity by modes in the 15-18 \mt range (where the 15.8, 16.4, 17.4 and 17.8 \mt PAH features reside; \citealt{ricca2010}). A small but perhaps important fraction of modes are predicted to emit between 18-20 \m. If we assume that the residual emission we observe in the 18-20 \mt zone is not a spurious result, then we must conclude that the carriers of this emission region are relatively insensitive to radiation field strength, as similar emission is observed in all four fields---despite varying PAH feature strengths and fine-structure line intensities between these fields. Thus, a possible carrier could be large, compact PAHs (c.f. the grandPAH hypothesis; \citealt{andrews2015}).

\subsection{PAH abundances}

We calculate the total far-infrared flux based on the modified blackbody fits to the SEDs. Combining this with the total PAH flux, we can use the formalism of \citet{tielensbook} to deduce the fraction of carbon locked in PAHs. We determine that the percentage of carbon locked in PAHs along our sight-lines toward C32 is approximately $2.9\pm0.4$\%, while C35 is a bit lower at $2.3\pm0.3$\%. These values are slightly depressed relative to values typical for the ISM ($\sim$3.5\%, \citealt{tielensbook}). In C32, this fraction is minimal at the ``knee" of the \ha channel at position C32-10 ($2.0\pm0.4$\%). The maximum for the C32 field occurs towards the far right at position C32-02 ($3.7\pm0.4$\%). No systematic variation across the C35 field is detected.

\subsection{Implications}

To first order, the dust and PAH emission properties are the same toward C32 and C35, despite being diametrically opposed across the Galactic center. This can be explained by the fact that C32 and C35 are both on boundaries of emission lobes (northern and southern, relative to the Galactic plane, respectively), which are thought to originate from a starburst roughly 7 Myr ago \citep{lutz1999,bland2003,law2010}. The dust grain temperatures ($\sim$20 K) in these environments are indistinguishable from those of diffuse cirrus grains. Similarly, the PAH band ratios measured toward cirrus clouds is similar but not identical to that observed in our bulge environments. Shock/turbulent heating is expected in these environments, consistent with the presence of the shock-tracing 25.89 \mt [O~\textsc{iv}] line.

When examined in detail, spatial variations within C32 and C35 are present. In C32 in particular, a region of strong \ha emission is present that is correlated with elevated fine-structure and [O~\textsc{iv}] line intensities and anticorrelated with PAH emission strength. This part of the C32 field may represent a transition zone for the outflow entering the surrounding medium.

The small variation seen in PAH characteristics in the C32 and C35 fields, despite significant variations in the fine-structure line emission, is perhaps akin to the similarity found by \citet{andrews2015} in the PDR peaks of 3 reflection nebulae. This may then be further evidence of the presence of a group of stable PAH molecules dominating the emission bands (i.e., grandPAHs; \citealt{andrews2015}). Moreover, the 15-20 \mt emission plateau exhibits underlying residual emission between 18-20 \mt that does not vary between all four fields we have examined, which may be additional evidence for a PAH population relatively insensitive to variations in radiation field strength.

\section{Conclusions}
\label{sec:conclusion}

We have analyzed \textit{Spitzer}/IRS spectra of diffuse emission toward the Galactic bulge. Combined with mid- and far-infrared photometry, we have investigated the spectral characteristics of our observations in the context of the local bulge environment. Our primary conclusions are as follows:

\begin{enumerate}

\item There is an evolution in spectral appearance with increasing distance from the Galactic Center. The spectra of C32 and C35 (located at (l, b) = (0.0\td, 1.0\td), (0.0\td, -1.0\td), respectively) are exceedingly similar, including strong PAH bands, fine-structure lines, plateaus, molecular emission and continuum emission. All of these features weaken at the position of the OGLE field, located at (l, b) = (0.4\td, -2.1\td). Its continuum emission is almost entirely dominated by zodiacal dust, though a weak rising continuum beyond 30 \mt is still present. At the most distant location, that of NGC 6522 (1.0\td, -3.8\td), the continuum emission is almost entirely due to zodiacal dust, and the PAH bands, plateaus and atomic lines are generally barely detectable.

\item The similarity of the C32 and C35 spectra may be explained by their locations: they both lie on boundaries of northern and southern outflow lobes, the former of which is known as the Galactic center lobe.

\item The PAH features in C35 are an almost exact match, in relative strength and profile shape, to the PAH features in the M82 superwind after removing their continua (at approximately 200\arcsec~from the center of M82, along its minor axis; region 1 of \citealt{beirao2015}). The C32 PAH features are similar to the M82 superwind but to a lesser extent. Thus, we have a local measurement of a galactic wind, which is common in star-forming galaxies such as M82.

\item Within our fields, the strength of the PAH and fine-structure features are related to a region of elevated \ha emission, which generally traces the lobe boundaries: generally, where the \ha emission is bright, the fine-structure lines are bright and the PAH bands are weak.

\item In contrast to the PAHs, the 25.89 \mt [O~\textsc{iv}] line peaks in/near the \ha channel in C32. This line is thought to be a tracer of shocked gas, confirming the presence of an outflow impacting the nascent medium---i.e., the Galactic center lobe. The [O~\textsc{iv}] line is also detected in the south (in C35), which is located within the more complex southern lobe environment. 

\item SED fitting indicates that the temperatures of thermal dust grains in C32 and C35 are $\sim$20 K, consistent with the temperature found for the diffuse ISM cirrus spectrum. These grains are expected to be heated not only radiatively but by shock and/or turbulent heating.

\item We infer that $2.9\pm0.4$\% and $2.3\pm0.3$\% of the total carbon along the sight-lines to C32 and C35 are locked in PAHs, respectively. This is somewhat less than typical ISM expectations of 3.5\%.

\item The 15-18 \mt emission plateau extends to 20 \mt in all four of our fields; the relative strength of the 17 \mt bump to underlying emission determines whether the plateau appears to only emit between 15-18 \mt versus 15-20 \m, if this is not a systematic error.

\item While distinct from the PAH sizes obtained towards supernova remnants, mean PAH sizes in C32 and C35 are comparable to those seen towards reflection nebulae, planetary nebulae and \HII~regions, and are thus typical. The average PAH size towards different positions in the superwind of M82 are somewhat smaller than those in C32 and C35.  

\end{enumerate}

The extreme similarity between the spectra towards C32 and C35, which are diametrically opposed from the Galactic center, is quite peculiar in some sense. Although they are on outflow boundaries, it is not immediately obvious why they are so alike. Not only are their dust continua comparable, but their plateaus, PAH bands, H$_2$ lines and 25.89 \mt [O~\textsc{iv}] emission strengths are also very similar. Only the atomic fine-structure lines appear to differ significantly between them. The PAH similarities may point to a small number of stable PAHs, grandPAHs, as dominating the PAH emission bands in these environments.

The natural questions to ask next are: are these sight-lines typical of the bulge environment? Is there a systematic dependence on Galactocentric distance? Will we find similar PAH, dust and line emission at other regions of the outflow boundaries? And finally, what is the relationship between emission towards the bulge and emission towards the general diffuse ISM?

Further detailed study of the exact variations within C32 may also be pertinent. We have focused on the position-to-position variations within the field, but careful analysis of the pixel-to-pixel variations may help us understand the link between \ha excitation, the possible release of Si and Fe from dust, and PAH excitation/destruction. The use of recent, higher spectral resolution multi-wavelength imaging (e.g., \ha) and comparisons to other extragalactic environments can help us answer these questions and probe starburst events and their influence on the PAH population(s).

\section*{Acknowledgments}

We thank Kris Sellgren for many helpful suggestions that improved the quality of this manuscript. We also thank William T. Reach for helping us estimate the zodiacal emission in our spectra. For kindly providing their data, we thank Pedro Beir{\~a}o, Adeline Caulet and Janet Simpson. The authors acknowledge support from NSERC discovery grants. MJS acknowledges support from a QEII-GSST scholarship. The IRS was a collaborative venture between Cornell University and Ball Aerospace Corporation funded by NASA through the Jet Propulsion Laboratory and Ames Research Center \citep{houck2004}. The Southern \ha Sky Survey Atlas (SHASSA) \citep{gaustad2001} was produced with support from the National Science Foundation. This research is based on observations with \textit{AKARI}, a JAXA project with the participation of ESA. This research has made use of NASA's Astrophysics Data System Bibliographic Services, and the SIMBAD database, operated at CDS, Strasbourg, France. This work has also made use of the Matplotlib Python plotting library \citep{hunter2007} and the Seaborn Python visualization library\footnote{\url{http://dx.doi.org/10.5281/zenodo.19108}}. This publication makes use of data products from the Two Micron All Sky Survey, which is a joint project of the University of Massachusetts and the Infrared Processing and Analysis Center/California Institute of Technology, funded by the National Aeronautics and Space Administration and the National Science Foundation.

\appendix
\renewcommand{\thefigure}{A\arabic{figure}}
\setcounter{figure}{0}
\renewcommand{\thetable}{A\arabic{table}}
\setcounter{table}{0}

\section{OGLE and NGC 6522 pointings}
\label{sec:pointings}

\begin{figure*}
\begin{center}
        \includegraphics[width=0.6\linewidth, clip=true, trim=2cm 1.2cm 2.2cm 1cm] {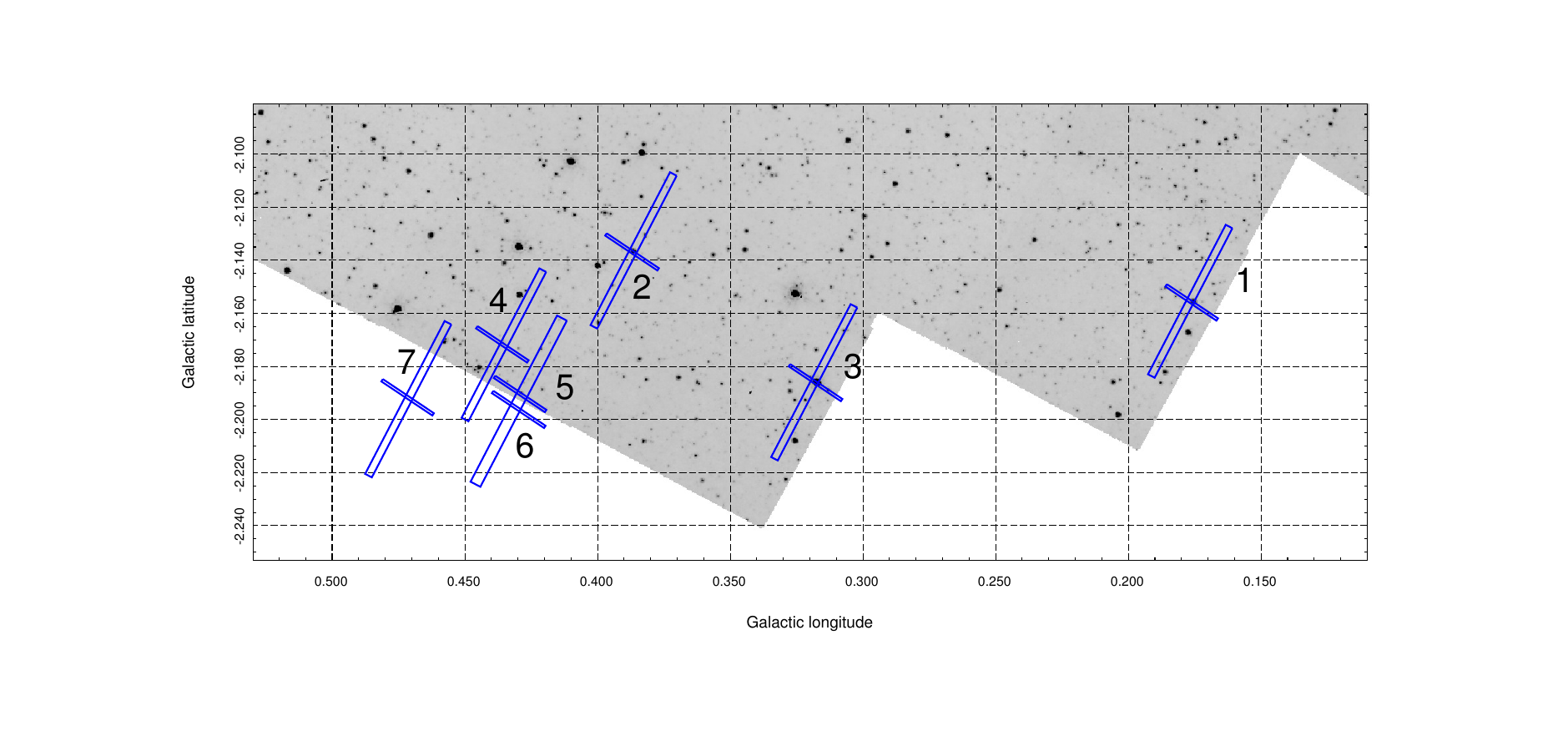}
\caption{The IRS apertures for OGLE overlaid on an IRAC 8 \mt image.The numbered labels correspond to the overlapping SL and LL apertures (the short and long blue rectangles, respectively); e.g., OGLE-3 (c.f. Table~\ref{table:obs}).}
\label{fig:irac_ogle}
\end{center}
\end{figure*}

\begin{figure*}
\begin{center}
        \includegraphics[width=0.5\linewidth, clip=true, trim=4.5cm 0.45cm 6.1cm 0.5cm] {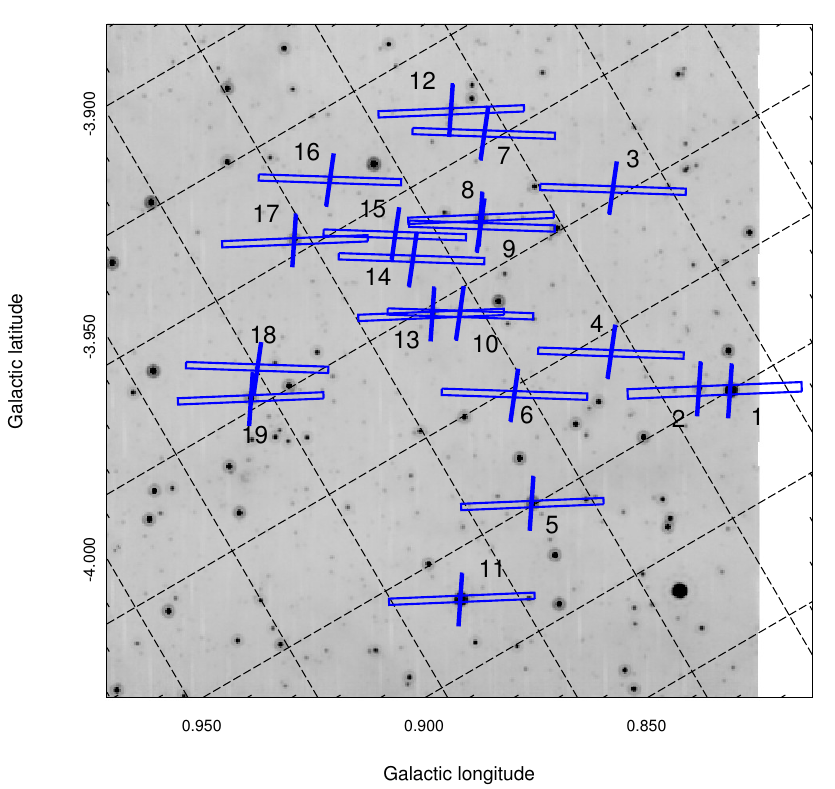}
\caption{The IRS apertures for NGC 6522 overlaid on a 24 \mt \textit{Spitzer} image acquired with the Multiband Imaging Photometer for Spitzer (MIPS; \citealt{rieke2004}). The numbered labels correspond to the overlapping SL and LL apertures (the short and long blue rectangles, respectively); e.g., NGC 6522-1 (c.f. Table~\ref{table:obs}).}
\label{fig:irac_ngc6522}
\end{center}
\end{figure*}

In Figs.~\ref{fig:irac_ogle} and~\ref{fig:irac_ngc6522} we present the pointings for the OGLE and NGC 6522 fields, respectively. The corresponding properties for these pointings are presented in Table~\ref{table:obs}.

\section{Band and line fluxes}
\label{sec:fluxes}

\begin{figure*}
\begin{center}
    \includegraphics[width=1\linewidth] {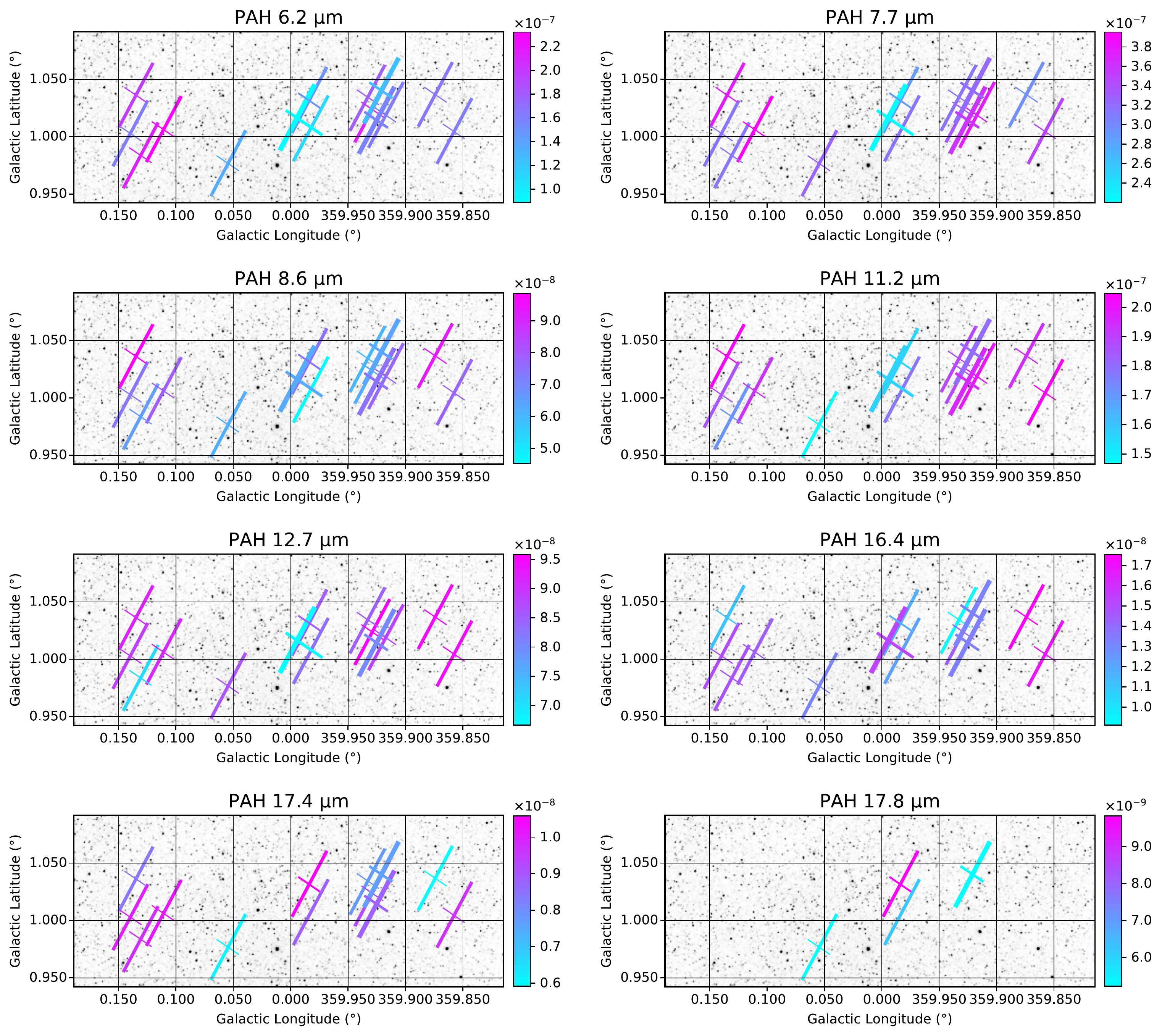}
\caption{The PAH emission band fluxes in the C32 fields, in Wm$^{-2}/$sr, overlaid on a 2MASS J-band image. A 3$\sigma$ signal-to-noise criterion has been applied to these data.}
\label{fig:c32maps}
\end{center}
\end{figure*}

\begin{figure*}
\begin{center}
    \includegraphics[width=1\linewidth] {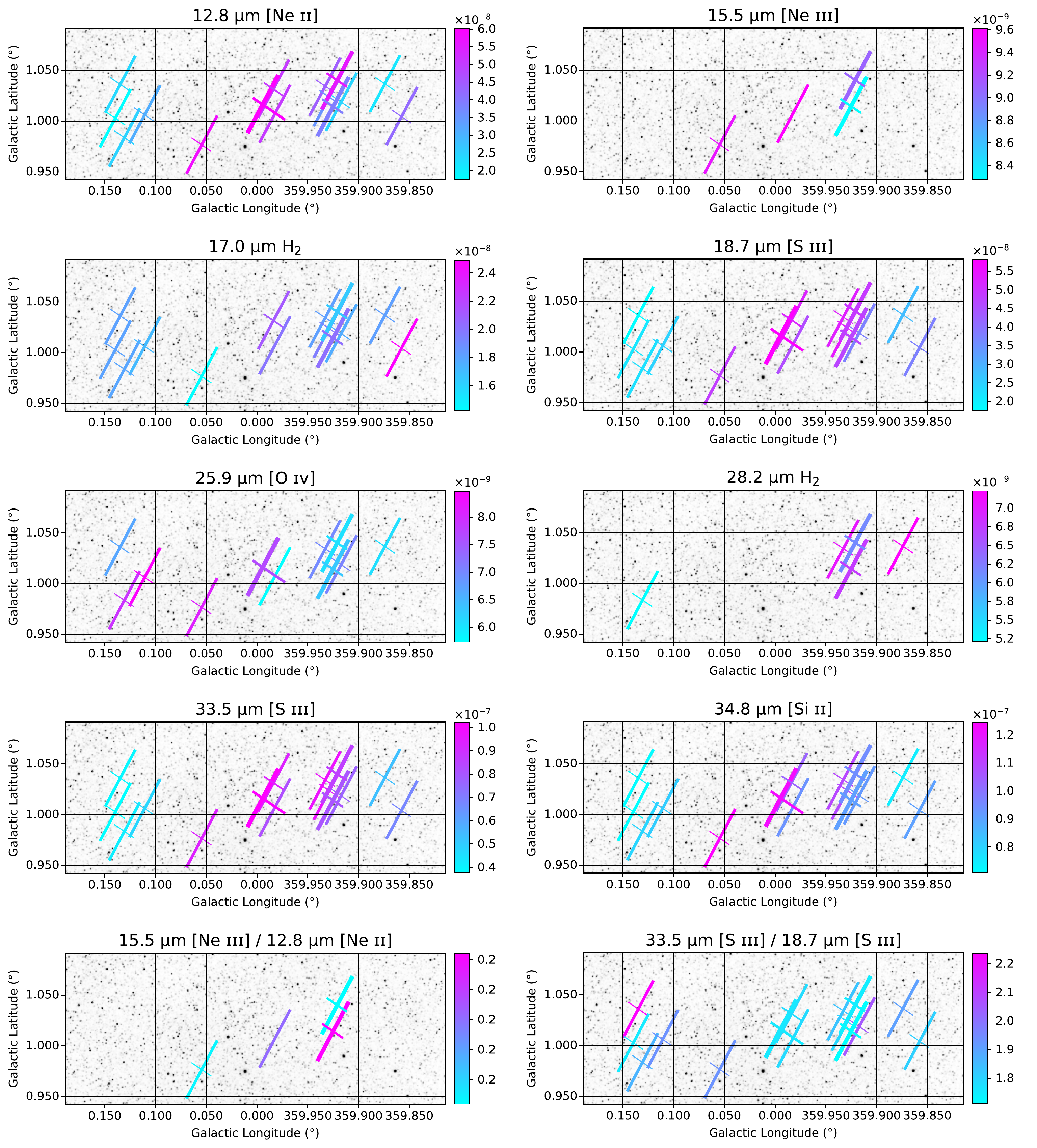}
\caption{The atomic and molecular emission line fluxes in the C32 fields, in Wm$^{-2}/$sr, overlaid on a 2MASS J-band image. A 3$\sigma$ signal-to-noise criterion has been applied to these data.}
\label{fig:c32maps2}
\end{center}
\end{figure*}

\begin{figure*}
\begin{center}
    \includegraphics[width=1\linewidth] {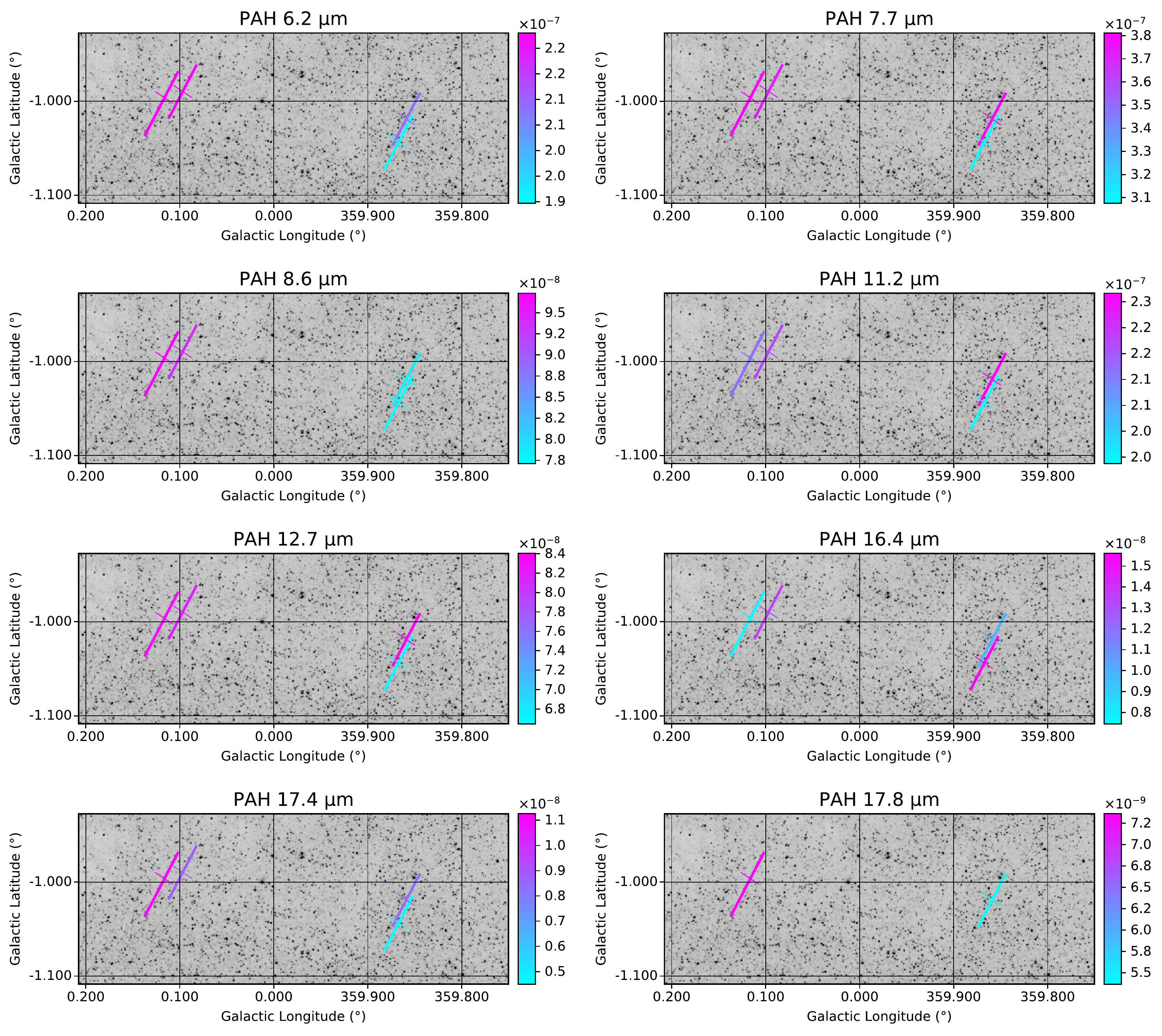}
\caption{The PAH emission band fluxes in the C35 fields, in Wm$^{-2}/$sr, overlaid on a 2MASS J-band image. A 3$\sigma$ signal-to-noise criterion has been applied to these data.}
\label{fig:c35maps}
\end{center}
\end{figure*}

\begin{figure*}
\begin{center}
    \includegraphics[width=1\linewidth] {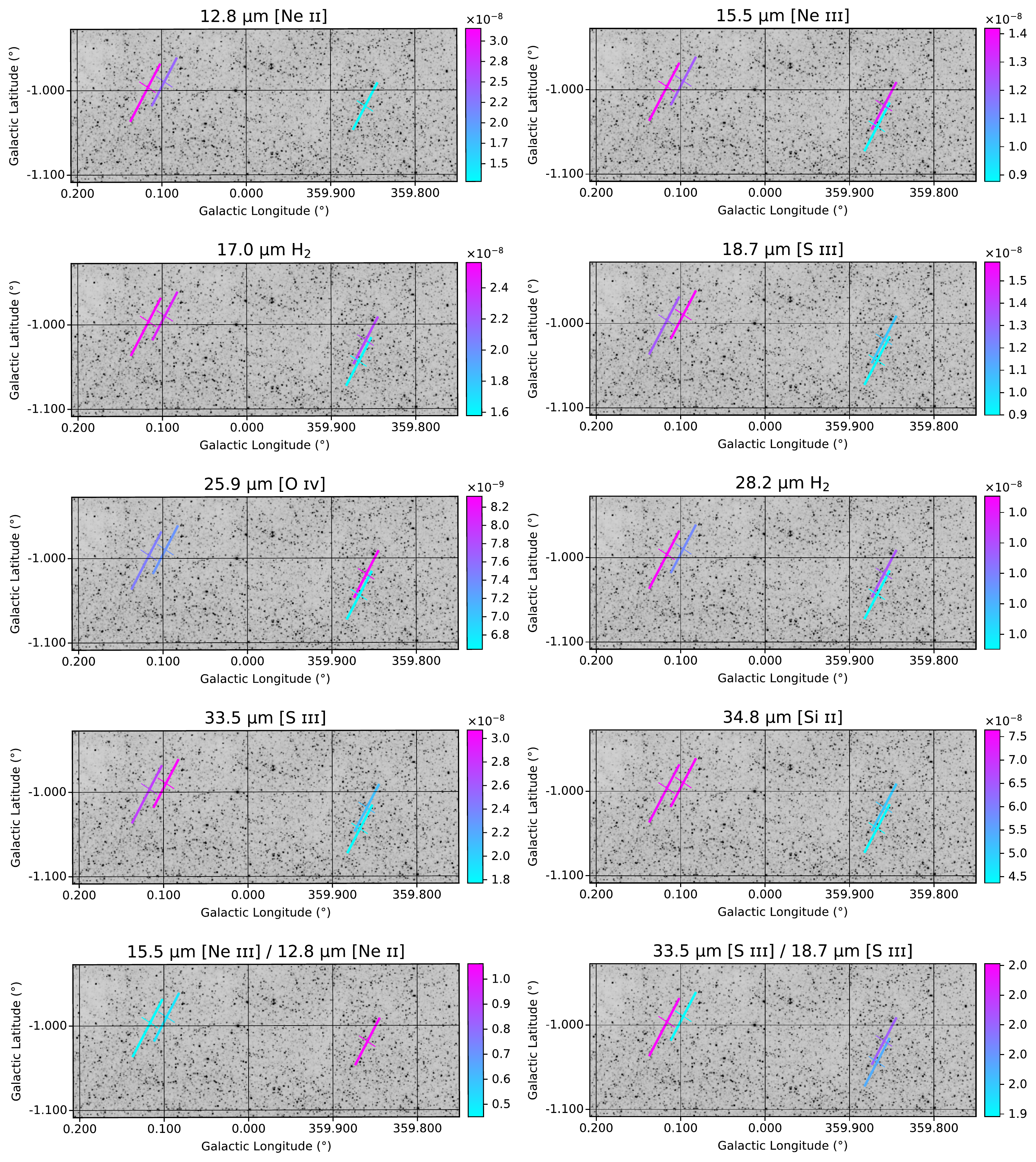}
\caption{The atomic and molecular emission line fluxes in the C35 fields, in Wm$^{-2}/$sr, overlaid on a 2MASS J-band image. A 3$\sigma$ signal-to-noise criterion has been applied to these data.}
\label{fig:c35maps2}
\end{center}
\end{figure*}

We include here tables of all measured quantities and maps of emission in C32 and C35 previously discussed in Section~\ref{sec:results}: Tables~\ref{table:pahfluxes} and~\ref{table:atomicfluxes} and Figs.~\ref{fig:c32maps} through~\ref{fig:c35maps2}, respectively.

\begin{table*}
    \centering
    \begin{tabular}{l*{7}rr}
    \toprule
    \toprule
	Object & \multicolumn{8}{c}{Feature} \\
	\cmidrule{2-9}
     & \thead{6.2 \m} & \thead{7.7 \m} & \thead{8.6 \m} & \thead{11.2 \m} &
    \thead{12.7 \m} & \thead{16.4 \m} & \thead{17.4 \m} & \thead{17.8 \m} \\
    \midrule

C32-1 & 16.5 (0.7) &  29.7 (0.9) &  9.3 (1.2) &  19.4 (1.2) &  9.5 (1.3) &  1.8 (0.2) &  0.6 (0.1) &  \\
C32-2 & 16.3 (0.5) &  34.8 (0.7) &  7.8 (0.8) &  20.5 (0.8) &  9.4 (0.9) &  1.7 (0.2) &  1.0 (0.2) &  \\
C32-3 & 12.5 (0.6) &  32.3 (0.8) &  6.4 (1.0) &  18.0 (1.0) &   &  1.3 (0.2) &  0.8 (0.1) &  0.5 (0.2) \\
C32-4 & 17.8 (0.5) &  31.7 (0.6) &  6.0 (0.8) &  18.9 (0.7) &  8.5 (0.9) &  0.9 (0.1) &  0.8 (0.1) &  \\
C32-6 & 16.2 (0.6) &  37.6 (0.8) &  7.7 (1.0) &  20.0 (1.0) &  8.8 (1.1) &   &   &  \\
C32-7 & 21.4 (0.7) &  33.4 (0.9) &  6.3 (1.1) &  18.8 (1.1) &  9.6 (1.2) &  1.4 (0.2) &  0.9 (0.2) &  \\
C32-8 & 15.8 (0.6) &  36.0 (0.7) &  7.5 (0.9) &  19.6 (0.9) &  8.0 (1.0) &  1.3 (0.2) &  0.9 (0.2) &  \\
C32-9 & 15.0 (0.7) &  28.9 (0.9) &  7.5 (1.2) &  15.6 (1.1) &  8.5 (1.3) &  1.2 (0.2) &  1.1 (0.1) &  1.0 (0.2) \\
C32-10 & 8.9 (0.8) &  22.0 (1.0) &  6.3 (1.3) &  15.7 (1.2) &  6.7 (1.4) &  1.5 (0.4) &   &  \\
C32-11 & 11.1 (0.6) &  31.5 (0.7) &  4.5 (0.9) &  17.8 (1.0) &  8.3 (1.1) &  1.2 (0.2) &  0.9 (0.2) &  0.6 (0.2) \\
C32-12 & 19.8 (0.4) &  37.5 (0.5) &  9.9 (0.7) &  20.4 (0.7) &  9.2 (0.8) &  1.1 (0.2) &  0.8 (0.2) &  \\
C32-13 & 13.6 (0.3) &  32.6 (0.4) &  6.2 (0.5) &  14.7 (0.5) &  8.6 (0.5) &  1.3 (0.2) &  0.6 (0.1) &  0.5 (0.2) \\
C32-14 & 23.2 (0.6) &  39.5 (0.7) &  7.9 (0.9) &  19.3 (0.9) &  9.1 (1.0) &  1.4 (0.3) &  1.0 (0.3) &  \\
C32-15 & 16.1 (0.6) &  31.4 (0.8) &  7.3 (1.0) &  18.6 (1.0) &  8.9 (1.1) &  1.5 (0.3) &  1.0 (0.2) &  \\
C32-16 & 21.3 (1.1) &  31.3 (1.4) &  6.6 (1.8) &  17.1 (1.8) &  7.1 (2.0) &  1.5 (0.4) &  1.0 (0.3) &  \\
C35-1 & 20.8 (0.7) &  37.9 (0.9) &  7.8 (1.1) &  22.7 (1.1) &  8.4 (1.2) &  1.0 (0.2) &  0.8 (0.1) &  0.5 (0.2) \\
C35-2 & 19.0 (1.2) &  30.8 (1.5) &  7.7 (2.0) &  19.4 (1.9) &  6.6 (2.2) &  1.6 (0.2) &  0.5 (0.1) &  \\
C35-3 & 22.3 (0.8) &  37.2 (1.0) &  9.4 (1.3) &  21.7 (1.2) &  8.2 (1.4) &  1.3 (0.2) &  0.9 (0.2) &  \\
C35-4 & 22.3 (1.0) &  38.1 (1.3) &  9.7 (1.7) &  21.3 (1.6) &  8.3 (1.9) &  0.7 (0.2) &  1.1 (0.2) &  0.7 (0.2) \\
C35-5 & 18.4 (1.3) &  32.8 (1.7) &  8.0 (2.1) &  21.1 (2.1) &  8.1 (2.4) &  1.5 (0.3) &  0.8 (0.2) &  \\
OGLE-1 & 7.4 (1.1) &  6.4 (1.4) &   &   &   &   &   &  \\
OGLE-2 &  &   &   &  5.8 (0.7) &   &   &   &  \\
OGLE-3 &  &   &   &  5.2 (1.1) &   &   &   &  \\
OGLE-4 & 2.1 (0.2) &  8.0 (0.3) &   &  5.2 (0.4) &  2.5 (0.4) &  0.5 (0.1) &   &  \\
OGLE-6 & 2.7 (0.3) &   &   &  5.0 (0.5) &  3.2 (0.6) &   &   &  \\
OGLE-7 &  &   &   &  4.6 (1.3) &   &   &   &  \\

    \bottomrule
    \end{tabular}
    \\[0.9ex]
    \parbox{0.9\linewidth}{\footnotesize \centering {\bf Notes.} In units of $10^{-8}$ Wm$^{-2}/$sr. Uncertainties are given in parentheses. A 3$\sigma$ signal-to-noise criterion is applied to our measurements; fields or positions with no bands meeting this criterion are omitted from the table.}
    \caption{PAH band fluxes}
    \label{table:pahfluxes}
\end{table*}

\begin{table*}
    \centering
    \begin{tabular}{l*{7}{rr}}
    \toprule
    \toprule
	Object & \multicolumn{8}{c}{Feature} \\
	\cmidrule{2-9}

     & \thead{[Ne~\textsc{ii}]}   & \thead{[Ne~\textsc{iii}]}  & \thead{H$_2$}  & \thead{[S~\textsc{iii}]}  & \thead{[O~\textsc{iv}]}   & \thead{H$_2$}  & \thead{[S~\textsc{iii}]}  & \thead{[Si~\textsc{ii}]} \\

     & \thead{12.8 \m}   & \thead{15.5 \m}  & \thead{17.0 \m}  & \thead{18.7 \m}  & \thead{25.9 \m}   & \thead{28.2 \m}  & \thead{33.5 \m}  & \thead{34.8 \m} \\
    \midrule

C32-1 & 2.2 (0.4) &   &  1.8 (0.4) &  2.8 (0.3) &  0.6 (0.2) &  0.7 (0.2) &  5.4 (0.3) &  7.4 (0.3) \\
C32-2 & 4.2 (0.3) &   &  2.5 (0.4) &  3.8 (0.4) &   &   &  6.8 (0.4) &  9.1 (0.5) \\
C32-3 & 5.6 (0.3) &  0.9 (0.2) &  1.6 (0.3) &  4.9 (0.3) &  0.6 (0.1) &  0.6 (0.1) &  8.6 (0.1) &  9.5 (0.2) \\
C32-4 & 4.4 (0.2) &   &  1.8 (0.5) &  5.3 (0.3) &  0.7 (0.2) &  0.7 (0.2) &  9.7 (0.3) &  11.0 (0.3) \\
C32-6 & 2.4 (0.3) &   &  1.8 (0.5) &  3.8 (0.5) &  0.7 (0.2) &   &  7.7 (0.4) &  9.2 (0.6) \\
C32-7 & 3.5 (0.3) &   &  2.0 (0.4) &  5.4 (0.4) &   &   &  9.7 (0.4) &  10.9 (0.5) \\
C32-8 & 3.9 (0.3) &  0.8 (0.2) &  2.0 (0.3) &  4.7 (0.3) &  0.6 (0.2) &  0.7 (0.2) &  8.1 (0.2) &  9.1 (0.3) \\
C32-9 & 5.0 (0.4) &   &  2.1 (0.3) &  5.2 (0.4) &   &   &  9.4 (0.4) &  10.4 (0.5) \\
C32-10 & 6.0 (0.4) &   &   &  5.8 (0.6) &  0.8 (0.2) &   &  10.2 (0.3) &  12.2 (0.6) \\
C32-11 & 5.0 (0.3) &  1.0 (0.3) &  2.0 (0.4) &  4.6 (0.4) &  0.6 (0.2) &   &  8.2 (0.3) &  9.4 (0.3) \\
C32-12 & 2.4 (0.2) &   &  1.8 (0.5) &  1.8 (0.4) &  0.7 (0.2) &   &  3.9 (0.4) &  7.2 (0.4) \\
C32-13 & 5.8 (0.1) &  0.9 (0.3) &  1.4 (0.5) &  4.8 (0.4) &  0.8 (0.2) &   &  9.4 (0.3) &  12.5 (0.4) \\
C32-14 & 3.1 (0.3) &   &  1.7 (0.5) &  2.5 (0.5) &  0.8 (0.3) &   &  4.8 (0.4) &  8.1 (0.5) \\
C32-15 & 1.8 (0.3) &   &  1.8 (0.5) &  2.1 (0.5) &   &   &  3.8 (0.4) &  7.1 (0.5) \\
C32-16 & 2.5 (0.6) &   &  1.8 (0.5) &  2.2 (0.6) &  0.8 (0.1) &  0.5 (0.1) &  4.2 (0.3) &  8.0 (0.3) \\
C35-1 & 1.3 (0.3) &  1.4 (0.2) &  2.3 (0.3) &  1.0 (0.2) &  0.8 (0.2) &  1.0 (0.2) &  2.1 (0.2) &  5.1 (0.2) \\
C35-2 &  &  0.9 (0.2) &  1.6 (0.2) &  0.9 (0.2) &  0.7 (0.1) &  1.0 (0.2) &  1.8 (0.2) &  4.4 (0.2) \\
C35-3 & 2.4 (0.4) &  1.2 (0.2) &  2.4 (0.4) &  1.6 (0.3) &  0.7 (0.1) &  1.0 (0.1) &  3.1 (0.1) &  7.6 (0.1) \\
C35-4 & 3.2 (0.5) &  1.4 (0.3) &  2.6 (0.3) &  1.3 (0.3) &  0.7 (0.2) &  1.1 (0.2) &  2.7 (0.2) &  7.5 (0.3) \\
C35-5 & 5.3 (0.7) &  1.2 (0.3) &  1.4 (0.4) &  4.2 (0.3) &  0.9 (0.1) &  0.7 (0.1) &  7.0 (0.2) &  13.4 (0.2) \\
OGLE-1 &  &   &   &   &   &   &   &  1.6 (0.5) \\
OGLE-2 &  &   &   &   &   &   &   &  1.6 (0.3) \\
OGLE-3 &  &   &   &   &   &   &   &  1.9 (0.3) \\
OGLE-4 & 0.3 (0.1) &   &   &   &   &   &   &  1.9 (0.3) \\
OGLE-6 &  &   &   &   &   &   &  0.6 (0.2) &  2.2 (0.2) \\
OGLE-7 &  &   &   &   &   &   &   &  1.7 (0.2) \\
NGC6522-5 &  &   &   &   &   &   &   &  0.9 (0.3) \\
NGC6522-6 &  &   &   &   &   &   &  0.4 (0.1) &  0.9 (0.1) \\
NGC6522-9 &  &   &   &   &   &   &   &  0.7 (0.2) \\
NGC6522-10 &  &   &   &   &   &   &   &  0.8 (0.2) \\
NGC6522-14 &  &   &   &   &   &   &   &  0.8 (0.2) \\
NGC6522-15 &  &   &   &   &   &   &   &  0.7 (0.2) \\
NGC6522-16 &  &   &   &   &   &   &   &  0.8 (0.2) \\
NGC6522-17 &  &   &   &   &   &   &   &  0.8 (0.3) \\
NGC6522-18 &  &   &   &   &   &   &   &  0.7 (0.2) \\

    \bottomrule
    \end{tabular}
    \\[0.9ex]
    \parbox{0.9\linewidth}{\footnotesize \centering {\bf Notes.} In units of $10^{-8}$ Wm$^{-2}/$sr. Uncertainties are given in parentheses. A 3$\sigma$ signal-to-noise criterion is applied to our measurements; fields or positions with no bands meeting this criterion are omitted from the table.}
    \caption{Atomic and molecular line fluxes}
    \label{table:atomicfluxes}
\end{table*}

\bibliographystyle{apj}

\end{document}